\documentclass[twocolumn,twocolappendix]{aastex631}
\usepackage{graphicx,natbib,bm,url,color}
\graphicspath{{./fig/}}

\newcommand{\ddt}{{\rm d}_\theta {}}

\newcommand{\DDr}{{\rm D}_r {}}

\newcommand{\DDt}{{\rm D}_\theta {}}

\newcommand{\dDt}{{\cal D}_\theta^2 {}}

\newcommand{\Drr}{{\rm D}_r^2 {}}

\def\qdiff{q_{\rm diff}}

\def\ggamma{\varsigma}

\newcommand{\EQ}{\begin{equation}}
\newcommand{\EN}{\end{equation}}
\newcommand{\EQA}{\begin{eqnarray}}
\newcommand{\ENA}{\end{eqnarray}}

\newcommand{\Eq}[1]{Equation~(\ref{#1})}
\newcommand{\Eqs}[2]{Equations~(\ref{#1}) and~(\ref{#2})}

\newcommand{\App}[1]{Appendix~\ref{#1}}

\newcommand{\Sec}[1]{Section~\ref{#1}}
\newcommand{\Secs}[2]{Sections~\ref{#1} and \ref{#2}}

\newcommand{\Fig}[1]{Figure~\ref{#1}}
\newcommand{\FFig}[1]{Figure~\ref{#1}}

\newcommand{\Tab}[1]{Table~\ref{#1}}

\newcommand{\bra}[1]{\langle #1\rangle}

\newcommand{\meanrho}{\overline{\rho}}

{}
{}

\newcommand{\meanSSSS}{\overline{\mbox{\boldmath ${\mathsf S}$}} {}}

{}
{}

{}
{}
{}
{}
{}
\newcommand{\meanEE}{\overline{\mbox{\boldmath $E$}}{}}{}
{}
{}
{}
{}
{}
{}
{}
{}
\newcommand{\meanUU}{\overline{\bm{U}}}

{}

\newcommand{\meanA}{\overline{A}}
\newcommand{\meanB}{\overline{B}}

\newcommand{\meanU}{\overline{U}}

\newcommand{\meanJ}{\overline{J}}

\newcommand{\meanp}{\overline{p}}

{}

{}
{}
{}

\newcommand{\pphi}{\hat{\bm{\phi}}}

\newcommand{\meanAA}{{\overline{\bm{A}}}}
\newcommand{\meanBB}{{\overline{\bm{B}}}}
\newcommand{\meanJJ}{{\overline{\bm{J}}}}

\newcommand{\kk}{\bm{k}}

\newcommand{\xx}{\bm{x}}

\newcommand{\rr}{\bm{r}}

\newcommand{\bb}{\bm{b}}
\newcommand{\BB}{\bm{B}}

\newcommand{\uu}{\bm{u}}

\newcommand{\nab}{{\bm{\nabla}}}

\newcommand{\ii}{{\rm i}}
\newcommand{\erf}{{\rm erf}}

\newcommand{\DD}{{\rm D} {}}

\newcommand{\dd}{{\rm d} {}}

\newcommand{\const}{{\rm const}  {}}

\def\H{\mbox{\rm H}}

\def\Rey{\mbox{\rm Re}}

\def\cs{c_{\rm s}}

\def\nuT{\nu_{\rm T}}
\def\etat{\eta_{\rm turb}}
\def\etaT{\eta_{\rm eff}}
\def\nuT{\nu_{\rm eff}}

\def\Beq{B_{\rm eq}}

\newcommand{\W}{\,{\rm W}}
\newcommand{\V}{\,{\rm V}}

\newcommand{\T}{\,{\rm T}}
\newcommand{\G}{\,{\rm G}}

\newcommand{\Jy}{\,{\rm Jy}}
\newcommand{\sr}{{\rm sr}}
\newcommand{\uJy}{\,{\mu\rm Jy}}

\newcommand{\MHz}{\,{\rm MHz}}

\newcommand{\K}{\,{\rm K}}
\newcommand{\g}{\,{\rm g}}
\newcommand{\s}{\,{\rm s}}

\newcommand{\uG}{\,\mu{\rm G}}

\newcommand{\cm}{\,{\rm cm}}

\newcommand{\m}{\,{\rm m}}

\newcommand{\kms}{\,{\rm km\,s}^{-1}}

\newcommand{\kpc}{\,{\rm kpc}}

\newcommand{\Gpc}{\,{\rm Gpc}}

\newcommand{\Gyr}{\,{\rm Gyr}}

\newcommand{\J}{\,{\rm J}}

\newcommand{\A}{\,{\rm A}}

\def\apjl{ApJL}
\def\apjs{ApJS}

\def\aap{A\&A}

\def\apjl{ApJL}
\def\apjs{ApJS}

\def\aap{A\&A}

\begin{document}

\title{Magnetic field spreading from stellar and galactic dynamos into the exterior}

\author[0000-0002-7304-021X]{Axel Brandenburg}
\affiliation{Nordita, KTH Royal Institute of Technology and Stockholm University, Hannes Alfv\'ens v\"ag 12, SE-10691 Stockholm, Sweden}
\affiliation{The Oskar Klein Centre, Department of Astronomy, Stockholm University, AlbaNova, SE-10691 Stockholm, Sweden}
\affiliation{McWilliams Center for Cosmology \& Department of Physics, Carnegie Mellon University, Pittsburgh, PA 15213, USA}
\affiliation{School of Natural Sciences and Medicine, Ilia State University, 3-5 Cholokashvili Avenue, 0194 Tbilisi, Georgia}

\author[0000-0003-2226-0025]{Oindrila Ghosh}
\affiliation{Oskar Klein Centre for Cosmoparticle Physics, Department of Physics, Stockholm University, AlbaNova, 10691 Stockholm, Sweden}

\author[0000-0002-2821-7928]{Franco Vazza}
\affiliation{Dipartimento di Fisica e Astronomia, Universita di Bologna, Via Gobetti 93/2, 40129 Bologna, Italy}
\affiliation{INAF Istituto di Radioastronomia, Via P. Gobetti 101, 40129 Bologna, Italy}

\author[0000-0002-6748-368X]{Andrii Neronov}
\affiliation{Universit\'e Paris Cit\'e, CNRS, Astroparticule et Cosmologie, 75006 Paris, France}
\affiliation{Laboratory of Astrophysics, \'Ecole Polytechnique F\'ed\'erale de Lausanne, 1015 Lausanne, Switzerland}

\begin{abstract}
The exteriors of stellar and galactic dynamos are usually modeled as current-free potential fields.
A more realistic description might instead be that of a force-free magnetic field.
Here, we suggest that, in the absence of outflows, neither of these reflect the actual behavior when the magnetic field
spreads diffusively into a more poorly conducting turbulent exterior outside dynamo.
In particular, we explain why the usual ordering, in which the dipole magnetic field is the most slowly decaying one, is altered,
and why the quadrupole can develop a toroidal component that decays even more slowly with radial distance.
This is a robust feature that persists even for spatially nonuniform magnetic diffusivities.
It is most clearly seen for spherical dynamo volumes and becomes more complicated for oblate ones.
In either case, however, these fields are confined within a magnetosphere, beyond which the field strength drops exponentially.
We demonstrate that the Faraday displacement current, which plays a role in a vacuum, can safely be neglected in all cases.
The superposition of magnetic fields from galaxies in the outskirts of voids between galaxy clusters
therefore cannot explain the magnetization of the intergalactic medium in voids,
reinforcing the conventional expectation that these fields are of primordial origin.
For quadrupolar configurations, the synchrotron emission from the magnetosphere is found to be constant along concentric rings.
The dipolar and quadrupolar configurations display large-scale radial trends that are potentially distinguishable with existing radio telescopes.
\end{abstract}

\section{Introduction}

The magnetic fields in stars and galaxies can be explained by dynamo
action, which converts the kinetic energy of turbulence and differential
rotation into magnetic energy.
Such systems are traditionally computed by assuming that the magnetic
field outside the dynamo domain continues as a current-free potential field.
Mathematically, this can be formulated as a suitable boundary condition applied
at the outer radius of a spherical \citep{SK69} or ellipsoidal domain \citep{Stix75}.
In cylindrical and several other coordinate systems, however, no method exists to apply a potential-field boundary condition.
One possibility is then to adopt a perfectly conducting boundary condition \citep{MER90}.
However, a more physical procedure may be to embed the galaxy in a poorly conducting exterior \citep{EMR90, BTK90},
as this would approximate a vacuum in the limit of zero conductivity.

The assumption of a vacuum outside the dynamo has never been thought to be more than a mathematically convenient construct.
However, in recent work by \cite{Garg+25}, a vacuum was assumed to exist even in the far field of galaxies.
Their study, which does not include outflows or winds of any type, led to subsequent investigations addressing the nature of
the magnetic field in the far field of galaxies \citep{Seller+Sigl2025, Ghosh+25}.
The present paper follows up on our earlier work \citep{Ghosh+25}.
Here, we focus on a more detailed numerical study, approach in some cases the limit of extremely poor conductivity,
and make contact with possible measurements of synchrotron emission.

It is commonly assumed that a more realistic boundary condition for a dynamo corresponds to an electrically conducting exterior,
though not necessarily to a perfectly conducting one.
In addition, if the exterior has low density and is at rest,
the magnetic field would be force-free, i.e., the cross product of current density and magnetic field vanishes.
In MHD, where the current density is proportional to the curl of the magnetic field, this implies that the two fields are parallel.
Such fields are therefore eigenfunctions of the curl operator and are
known as Chandrasekhar--Kendall functions \citep[after the early paper by][]{CK57}.
In the context of plasma confinement configurations, force-free magnetic fields are likewise often more realistic
than the assumption of a vacuum \citep{Freidberg14}.

Adopting a perfectly force-free magnetic field is generally problematic, because such configurations may not be topologically realizable
and would take significant time to establish in the vast exterior of a dynamo.
To alleviate this problem, a commonly used approach in solar physics is the magneto-frictional method \citep{yang1986force},
in which the Lorentz force drives flows through a friction term.
These flows relax the field until the force is minimized.
In a similar approach, we study here the spreading of galactic magnetic fields into intergalactic space by solving the
mean-field momentum equation and modeling the dynamo exterior as a medium with a certain viscosity and electrical conductivity.
A vacuum corresponds to the limit of very low conductivity, or equivalently very high magnetic diffusivity.
A large effective magnetic diffusivity can emerge in a turbulent environment when modeling the large-scale field as an average or mean field, $\meanBB$,
where the overbar denotes an appropriate averaging procedure.
However, as we explain below, the diffusivity needed to reproduce the electrical properties of a vacuum would be at least 8 orders of magnitude larger
than can reasonably be explained by turbulence.

It is useful to represent the resulting magnetic field as a multipole expansion with terms proportional to spherical harmonics
$Y_\ell^m(\theta,\phi)$, where $\ell$ and $m$ are the spherical harmonic degree and order, respectively.
For a vacuum, the magnetic field decays with radius $r$ as $r^{-(\ell+2)}$.
The lowest multiple is the dipole with $\ell=1$, for which the field decays
as $r^{-3}$, while for a quadrupole, $\ell=2$, it decays as $r^{-4}$.
In an electrically conducting environment, magnetic fields are expected to be close to force-free and can therefore decay differently.
If the gas pressure is negligible, force-free magnetic fields obey $\meanJJ\times\meanBB=0$, i.e.,
the current density $\meanJJ=\nab\times\meanBB/\mu_0$ with $\mu_0$ being the vacuum permeability, is parallel to $\meanBB$.
This means that $\nab\times\meanBB=\alpha_\mathrm{ff}\meanBB$, where
$\alpha_\mathrm{ff}$ is a coefficient that is constant along magnetic field lines, but may vary perpendicular to them \citep{Priest}.

To explain the toroidal magnetic fields observed around solar-type stars, \cite{Bonanno16} and \cite{Bonanno+DelSordo17}
considered force-free magnetic fields in mean-field dynamo exteriors.
They assumed the coronal magnetic field of those stars to be harmonic, but
did not address how such configurations can be achieved on finite timescales.
Here, we show that such fields can only be produced within a finite but slowly expanding magnetosphere around the dynamo.
This also has implications for galactic dynamos and for the question whether such configurations can explain the magnetic fields
in the voids between galaxy clusters \citep{Garg+25, Seller+Sigl2025, Ghosh+25}.

\cite{Ghosh+25} showed that the growth of the magnetosphere changes from ballistic to diffusive once the dynamo saturates.
They also demonstrated that a quadrupolar magnetic field in the magnetosphere decays with radius $r$ only as $r^{-2}$,
which is slower than the $r^{-3}$ decay of a dipolar field---both in a vacuum and in a conducting exterior.
Here, we present a detailed analysis of these empirical findings and
elucidate the origin of the slow radial decay for quadrupolar fields.
In particular, we explore the vacuum limit considered by \cite{Garg+25},
and compute the resulting synchrotron emission from our quadrupolar field.

In \Sec{Model}, we present the details of our model, allowing for
different profiles of the effective magnetic diffusivity, as well as
feedback from mean flows and, in some cases, the effects of a finite speed of light.
Our results on the establishment of a time-dependent magnetosphere are presented in \Sec{Results}.
In \Sec{Observational}, we discuss new observational constraints in terms of synchrotron radio emission,
which go beyond the rotation measure diagnostics presented in \cite{Ghosh+25}.
Finally, we conclude in \Sec{Conclusions}.

\section{Our model}
\label{Model}

\subsection{Basic equations}

We solve the mean-field dynamo equations using spherical coordinates, $(r,\theta,\phi)$.
Before making the MHD approximation, in which the Faraday displacement current is omitted,
we begin with the full Maxwell equations, which are also applicable to
an extremely poorly conducting medium or the vacuum considered by
\cite{Garg+25}.
To quantify the difference between a poorly conducting exterior and a vacuum,
we also include in some cases the Faraday displacement current.
The detailed motivation for this approach, along with first results, was presented by \cite{Ghosh+25}.

The displacement current can become important when the ordinary current density is small.
We assume that the mean current density $\meanJJ$ obeys Ohm's law, which in our case takes the form
\begin{equation}
\meanJJ=\sigma(\meanEE+\meanUU\times\meanBB+\overline{\uu\times\bb}),
\label{OhmsLaw0}
\end{equation}
where $\sigma$ is the microphysical conductivity, $\meanEE$ is the mean electric field, $\meanUU$ is the mean velocity, $\meanBB$ is the mean magnetic field,
and $\overline{\uu\times\bb}$ is the mean electromotive force resulting from the small-scale velocity and magnetic fields, $\uu$ and $\bb$, respectively.
In its simplest form, the mean electromotive force can be written as
\begin{equation}
\overline{\uu\times\bb}=\alpha\meanBB-\etat\,\mu_0\meanJJ,
\label{OhmsLaw1}
\end{equation}
where $\alpha$ is a coefficient that emerges in the calculation of $\overline{\uu\times\bb}$ for helical turbulence \citep{Mof78}.
It can lead to the growth and sustenance of a mean magnetic field and is generally referred to as the $\alpha$-effect.
Furthermore, the quantity $\etat$ denotes the turbulent magnetic diffusivity.

It is convenient to define the microphysical magnetic diffusivity $\eta=(\mu_0\sigma)^{-1}$ along with the total (effective) magnetic diffusivity $\etaT=\eta+\etat$.
Ohm's law for the mean current density can then be written as
\begin{equation}
\mu_0\meanJJ=(\meanEE+\meanUU\times\meanBB+\alpha\meanBB)/\etaT.
\label{OhmsLaw2}
\end{equation}
The term $\mu_0\meanJJ$ appears as the second term on the right-hand side (rhs)
of the averaged Amp\`ere--Maxwell equation,
\begin{equation}
\frac{1}{c^2}\frac{\partial\meanEE}{\partial t}
=\nab\times\meanBB-\mu_0\meanJJ.
\label{dE1dt}
\end{equation}
where $c$ is the speed of light.

The full set of Maxwell equations can then be written as
\begin{equation}
\frac{\partial\meanAA}{\partial t}=-\meanEE,\qquad
\meanBB=\nab\times\meanAA,
\label{dAdt}
\end{equation}
along with
\begin{equation}
-\left(1+\frac{\etaT}{c^2}\frac{\partial}{\partial t}\right)\meanEE
=\meanUU\times\meanBB+\alpha\meanBB-\etaT\nab\times\meanBB,
\label{dE2dt}
\end{equation}
where $\meanAA$ is the mean magnetic vector potential, so that the mean magnetic field is given by $\meanBB=\nab\times\meanAA$, which satisfies the constraint $\nab\cdot\meanBB=0$.
The mean charge density is given by $\meanrho_\mathrm{e}=\epsilon_0\nab\cdot\meanEE$, where $\epsilon_0=1/\mu_0 c^2$ is the vacuum permittivity.
Here, we have adopted the Weyl gauge, i.e., the electrostatic potential vanishes.

The Faraday displacement current, $\epsilon_0\partial\meanEE/\partial t$, can be neglected when $\etaT\ll\tau c^2$,
where $\tau$ is a characteristic time scale.
In this limit, the MHD approximation applies and the terms on the rhs of \Eq{dE2dt} can be directly inserted
into the rhs of \Eq{dAdt}, yielding the familiar induction equation.
We refer to \App{DispersionRelation} for a discussion of the dispersion relation
of a dynamo with an $\alpha$-effect and a finite value of $c$ for the simple Cartesian case in a periodic domain
with constant coefficients and no feedback from mean gas motions, i.e., $\meanUU=0$.
There, and throughout the main paper, we identify $\tau$ with the light
travel time, i.e., with $(ck)^{-1}$ or $R/c$, where $k$ is the wavenumber of the
lowest dynamo mode in a periodic domain, and $R$ is the radius of the
dynamo in an open domain.
Thus, we have $\etaT\ll c/k$ and $\etaT\ll cR$ in these two cases.

It is important to stress that the nature of the basic equations changes from parabolic to hyperbolic when the displacement current is included.
This also implies that the computational time-step constraint changes from being quadratic to linear in the mesh spacing,
which can significantly alleviate the computational restrictions of explicit schemes.

To account for the mean gas motions driven by the magnetic field, we also solve in some cases the momentum equation, assuming an isothermal gas.
The sound speed $\cs$ is then constant, and the mean pressure $\meanp(x,z,t)$ is given by $\meanp=\meanrho\cs^2$.
Mean-field hydromagnetic equations of this type have been studied by many authors, starting with the work by \cite{Schuessler79} in Cartesian geometry,
and subsequently \cite{Bra+92}, \cite{Kitchatinov+Ruediger95}, and \cite{Rempel06}.
The full system of equations is then complemented by those for $\meanrho$ and $\meanUU$, i.e.,
\begin{equation}
\frac{\DD\ln\meanrho}{\DD t}=-\nab\cdot\meanUU,
\label{dlnrhodt}
\end{equation}
\begin{equation}
\frac{\DD\meanUU}{\DD t}=-\cs^2\nab\ln\meanrho
+\left[\meanJJ\times\meanBB+\nab\cdot(2\nuT\meanrho\meanSSSS)\right]/\meanrho,
\label{dUdt}
\end{equation}
where $\DD/\DD t=\partial/\partial t+\meanUU\cdot\nab$ is the advective derivative,
and $\meanSSSS$ is the rate-of-strain tensor of the mean flow, with
components $\overline{\mathsf{S}}_{ij}=(\meanU_{i;j}+\meanU_{j;i})/2
-\delta_{ij}\nab\bm\cdot\meanUU/3$.
Here, semicolons denote covariant derivatives.

There are three mean-field parameters: the $\alpha$-effect, the turbulent (effective) viscosity $\nuT$,
and the turbulent (effective) magnetic diffusivity $\etaT$.
In addition to the macrophysical backreaction from the mean Lorentz force $\meanJJ\times\meanBB$, we also allow for microphysical feedback in the
form of $\alpha$-quenching \citep{IR77}, and assume $\alpha$ to be proportional to a quenching factor,
\begin{equation}
Q(\meanBB)=1/(1+\meanBB^2/\Beq^2),
\end{equation}
where $\Beq$ is the equipartition field strength above which $\alpha$
begins to be affected by the Lorentz-force feedback from the small-scale magnetic field.

\subsection{Details of the model}
\label{Details}

In the following, we adopt the computational domain $r_\mathrm{in}\leq r\leq r_\mathrm{out}$ and $0\leq\theta\leq\pi/2$,
and assume axisymmetry, i.e., $\partial/\partial\phi=0$.
In \Eq{OhmsLaw2}, we must specify spatial profiles for $\alpha$ and $\etaT$.
To model a localized dynamo-active region, we choose $\alpha$ to be finite around the origin and zero outside the dynamo region.

We mostly study spherical dynamo regions with a given radius $R$, but in some cases we also allow the region to be oblate.
In those cases, the disk thickness in the $z$ direction, i.e., along the axis, is denoted by $h$.
To model a smooth transition, we adopt an error-function profile.
In addition, $\alpha$ should be antisymmetric about $z=0$.
This is accomplished by multiplying by an additional factor $z/h$.
Thus, we write the $\alpha$-effect in the form
\begin{equation}
\alpha=\alpha_0 \frac{z}{h}\,
\frac{Q(\meanBB)}{2}\left[1-\erf\frac{\sqrt{(\varpi/R)^2+(z/h)^2}-1}{w/R}\right],
\end{equation}
where $\alpha_0$ is a constant, $\varpi=r\sin\theta$ is the cylindrical radius,
$z=r\cos\theta$ is the height above the midplane, and $w$ is the width
of the transition from the dynamo-active region to the exterior.
For $h=R$, the dynamo region is spherical.

For $\etaT$, we allow the effective magnetic diffusivity to take a larger $\etaT^\mathrm{ext}$ value
outside the dynamo region and equal to $\etaT^\mathrm{int}\leq\etaT^\mathrm{ext}$ within it.
We construct the profile for $\etaT$ in a manner similar to that of $\alpha$, except that it is symmetric about $z=0$
and does not include a quenching factor.
Thus, we choose the profile for $\etaT$ to be given by
\begin{equation}
\etaT=\etaT^\mathrm{int}+\frac{\etaT^\mathrm{ext}-\etaT^\mathrm{int}}{2}
\left[1+\erf\frac{\sqrt{(\varpi/R)^2+(z/h)^2}-1}{w/R}\right].
\end{equation}
We also study some models, in which $\etaT$ is uniform throughout the domain, i.e., $\etaT^\mathrm{ext}=\etaT^\mathrm{int}$,
as well as models in which $\etaT$ increases gradually in the exterior following a power law.
In those cases, we take
\begin{equation}
\etaT^\mathrm{ext}=\etaT^\mathrm{int}\left(1+r/r_0\right)^m,
\label{power_eta}
\end{equation}
where $m>0$ is the exponent and $r_0$ is a parameter that determines how rapidly $\etaT$ increases in the exterior.

To avoid the coordinate singularity at $r=0$, we assume a finite value of $r_\mathrm{in}=0.1\,R$ in most cases.
For the outer radius, we assume $r_\mathrm{out}=1000\,R$.
We restrict ourselves to solving the equations in the first quadrant of the meridional plane, i.e., $0\leq\theta\leq\pi/2$.
This means that, for dipolar fields, $\meanB_\theta$ is finite at the equator, while $\meanB_r$ and $\meanB_\phi$ are vanishing.
For quadrupolar fields, by contrast, $\meanB_r$ and $\meanB_\phi$ are finite at equator, while $\meanB_\theta$ is vanishing.
In terms of the components of $\meanAA$, we thus impose
the following boundary conditions on $\theta=\pi/2$ \citep{Jabbari+15}:
\begin{equation}
\frac{\partial\meanA_r}{\partial\theta}=\meanA_\theta=
\frac{\partial\meanA_\phi}{\partial\theta}=0
\end{equation}
for dipolar (odd-parity) fields, and
\begin{equation}
\meanA_r=\frac{\partial\meanA_\theta}{\partial\theta}=\meanA_\phi=0
\end{equation}
for quadrupolar (even-parity) fields.
In the radial direction, we impose a perfect-conductor boundary condition at both ends, i.e., $\meanA_\theta=\meanA_\phi=0$.
Owing to the use of the Weyl gauge, analogous boundary conditions in both directions also apply to the components of $\meanEE$.

In all our models, we use as an initial condition weak, Gaussian-distributed noise for $\meanAA$.
This acts as a seed magnetic field required for the dynamo to amplify the field exponentially once $\alpha_0$ is large enough.
Owing to the $Q(\meanBB)$ factor, the growth becomes quenched and $|\meanBB|$ reaches values close to $\Beq$.
The initial values of $\meanEE$ and $\meanUU$ are set to zero, and $\meanrho$ is set to a constant reference value $\rho_0$.
We recall that gravity is not included in our models.

The strength of the $\alpha$-effect is quantified by the dynamo number $C_\alpha=\alpha_0 R/\etaT^\mathrm{int}$.
The diffusivity contrast is given by $C_\eta=\etaT^\mathrm{ext}/\etaT^\mathrm{int}$.
Time is often expressed in terms of the diffusion time $\tau_\mathrm{diff}=R^2/\etaT^\mathrm{ext}$ based on the exterior diffusivity.
During the kinematic growth phase, the growth rate $\gamma$ of the magnetic field is characterized
by the nondimensional number $C_\gamma=\gamma R^2/\etaT^\mathrm{int}$.
Unlike $C_\alpha$ and $C_\eta$, which are input parameters, $C_\gamma$ is an output parameter.

For all of our spherical simulations, we use a mesh of $8192\times32$ points in $r_\mathrm{in}\leq r\leq1000\,R$ and $0\leq\theta\leq\pi/2$.
We use the \textsc{Pencil Code} \citep{PC}, where the inclusion of the displacement current is facilitated by the
\texttt{special/disp\_current.f90} module, and
its mean-field implementation through the \texttt{magnetic/meanfield.f90} module.
This implementation has been validated by reproducing $\gamma(k)$ from the dispersion relation for constant coefficients
in the Cartesian case; see \App{DispersionRelation}.
For earlier work with the \textsc{Pencil Code} in spherical geometry, where the mean-field momentum equation was included,
we refer to the papers by \cite{Jabbari+15} and \cite{JB21}.
In both studies, gravity played an important role; however, in the present case, it could lead to the development of a Parker wind,
which we want to suppress.
Therefore, we set here the gravitational potential to zero.

As an empirical measure of the front speed in simulations including the displacement current,
we determine the radius $r_{0.01}$ where the compensated value of the latitudinally averaged magnetic field
has dropped to 0.01 times the value inside the magnetosphere.
We expect that $r_{0.01}/ct<1$ in all cases, which is indeed borne out by the simulations.

As already discussed in \cite{Ghosh+25}, the front speed is constant during the kinematic growth phase.
We refer to this growth as ballistic.
In the saturated state, the front speed decreases as the radius grows diffusively.
Thus, for the front radius $r_\ast(t)$, we have 
\begin{equation}
r_\ast^2(t)=\left\{
\begin{array}{ll}
q_\mathrm{ballistic} \, \gamma\etaT^\mathrm{ext} \, (t-t_\ast)^2 & \mbox{(kinematic phase),}\\
q_\mathrm{diff} \, \etaT^\mathrm{ext} \, (t-t_\ast) & \mbox{(saturated phase),}
\end{array}
\right.
\label{FrontRadius}
\end{equation}
where $q_\mathrm{ballistic}$ and $q_\mathrm{diff}$ are empirical coefficients of order unity,
and $t_\ast$ denotes the time at which the corresponding growth behavior becomes fully established.

\subsection{Application to galactic scales}
\label{ApplicationGalaxies}

In this work, we present our results in dimensionless form by expressing length in units of $R$
and time in units of $\tau_\mathrm{diff}\equiv R^2/\etaT^\mathrm{ext}$.
In a specific example, \cite{Ghosh+25} estimated $\etaT^\mathrm{ext}=3000\kpc\kms$.
They presented their results in dimensional form.
The corresponding dimensionless time at the end of their runs, $t_\mathrm{end}/\tau_\mathrm{diff}$, was about 1400.
Identifying $t_\mathrm{end}$ with the Hubble time of $14\Gyr$ yields $\tau_\mathrm{diff}=0.01\Gyr$.
The length unit is therefore $R=(\etaT^\mathrm{ext}\tau_\mathrm{diff})^{1/2}\approx5.5\kpc$.
For one Hubble time, \cite{Ghosh+25} quoted a typical diffusion length of $200\kpc$; see their Table~II
for a case they refer to as ``cosmic web dynamics.''

\begin{table*}\caption{
Summary of runs presented in the paper.
}\hspace{-10mm}\vspace{12pt}\centerline{\begin{tabular}{cccccccccccccc}
Run & sym & $\meanU$ & $cR/\etaT$ & $r_\mathrm{in}/R$ & $h/R$ & $m$ & $C_\alpha$ & $C_\eta$ & $C_\gamma$ & $B_0/\Beq$ &
$\qdiff$\tablenotemark{a} & $r_{0.01}/ct$\tablenotemark{b} & $n$ \\
\hline
A & D & $\neq0$ & $\infty$ & 0.1 &  1  & 0 & 25 & 50 &  63  & 0.7 & 4.9 & --- & 3 \\
B & Q & $\neq0$ & $\infty$ & 0.1 &  1  & 0 & 25 & 50 &  66  & 0.4 & 2.5 & --- & 2 \\
C & Q & $\neq0$ & $\infty$ & 0.1 &  1  & 0 & 25 & 10 &  70  & 0.4 & 2.5 & --- & 2 \\
D & D &   $0$   & $\infty$ & 0.1 &  1  & 0 & 25 &  1 &  74  & 0.7 & 4.9 & --- & 3 \\
E & D &   $0$   &    1     & 0.1 &  1  & 0 & 25 &  1 &  7.2 & 0.7 & --- & 0.8 & 3 \\
F & D &   $0$   &   0.5    & 0.1 &  1  & 0 & 25 &  1 &  3.7 & 0.7 & --- & 0.6 & 3 \\
G & Q &   $0$   & $\infty$ & 0.1 &  1  & 0 & 25 &  1 &  77  & 0.4 & 2.5 & --- & 2 \\
G'& Q &   $0$   & $\infty$ & 0.1 &  1  & 1 & 25&$r^{-m}$& 68  & 0.4 & 2.5 & ---&2.16\\
G''&Q &   $0$   & $\infty$ & 0.1 &  1  &1.5& 25&$r^{-m}$& 68  & 0.4 & 2.5 & ---&2.23\\
H & Q &   $0$   &    1     & 0.1 &  1  & 0 & 25 &  1 &  8.2 & 0.4 & --- & 0.7 & 2 \\
I & Q &   $0$   &   0.5    & 0.1 &  1  & 0 & 25 &  1 &  4.2 & 0.4 & --- & 0.6 & 2 \\
J & Q &   $0$   & $\infty$ & 0.2 &  1  & 0 & 50 &  1 & 386  & 0.6 & 2.5 & --- & 2 \\
K & Q &   $0$   & $\infty$ & 0.1 & 0.5 & 0 & 50 &  1 & 348  & 0.2 & 2.5 & --- & 2 \\
L & Q &   $0$   & $\infty$ & 0.1 & 0.2 & 0 & 50&1&78&$2\times10^{-5}$&2.5&--- & 2 \\
\label{TSummary}\end{tabular}}
\tablenotetext{a}{A hyphen indicates that \Eq{eq:Bfit} does not provide an appropriate fit.}
\tablenotetext{b}{The ratio $r_{0.01}/ct$ can only be evaluated when $c$ is finite.}
\end{table*}

We reiterate that our intention is not to provide a realistic model of a galactic dynamo.
Instead, we aim to understand the far-field behavior of a generic dynamo.
Our results may therefore in principle also be applicable to stellar dynamos.
We stress that differential rotation is not included either.
All these simplifications are made to keep the system sufficiently transparent
to identify potentially generic behaviors
that are expected to occur in the far field of astrophysical dynamos in general.

\section{Results}
\label{Results}

\subsection{Exponential growth phase}

Let us first look at the initial spreading of a dynamo-generated magnetic field into the region $r>R$.
We adopt an $\alpha$-effect in the range $r_\mathrm{in}\leq r\leq R$ with $C_\alpha=25$,
and assume a 10- or 50-fold increase of $\etaT$ in the exterior;
see \Tab{TSummary} for a summary of the runs discussed in this paper.
The three runs of \cite{Ghosh+25} are included in \Tab{TSummary} as Runs~A--C.

At early times, the magnetic field expands into $r>R$ and is found to fall off with distance as $e^{-\kappa r}$, where
\begin{equation}
\kappa=(\gamma/\etat)^{1/2}
\label{Kappa}
\end{equation}
is the radial decay rate, $\gamma$ is the growth rate, and $\etat$ the turbulent magnetic diffusivity.
This scaling is analogous to that for the skin effect, except that here
we use an imaginary frequency, which leads to a nonoscillatory decay with increasing distance from the dynamo regime.

\begin{figure}\begin{center}
\includegraphics[width=\columnwidth]{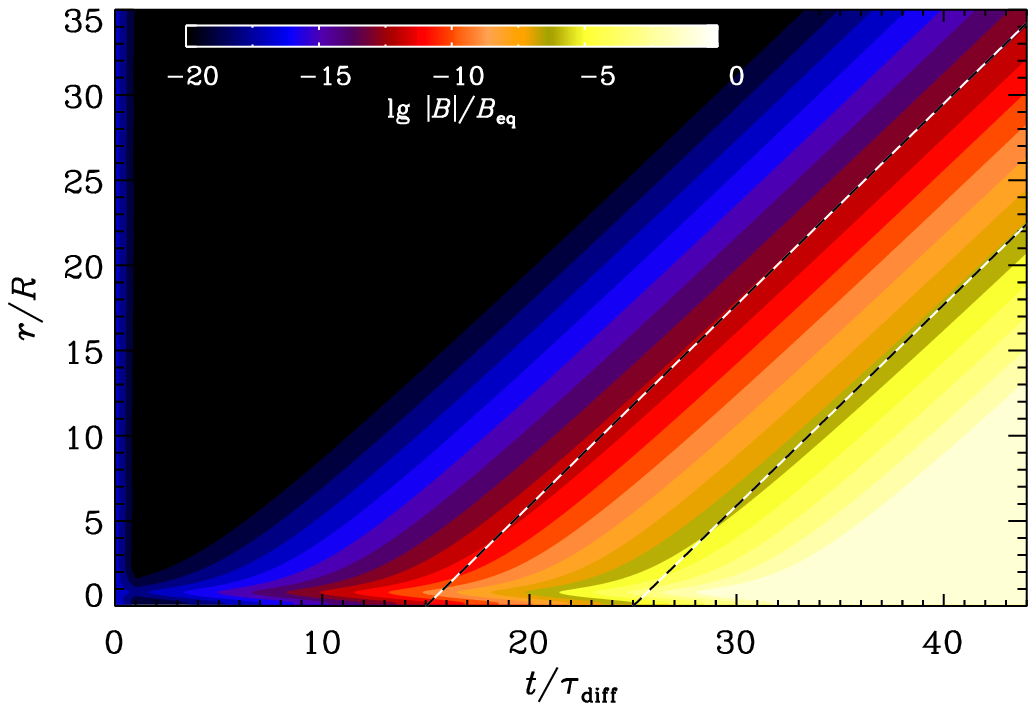}
\end{center}\caption{
Color-scale representation of $\ln|\BB|$ vs.\ $r$ and $t$ for Run~B.
Yellow (blue) shades denote large (small) field strengths.
The dynamo operates in $0\leq r\leq 1$, as can be seen by
the elevated field strength close to $r=0$.
The left (right) dashed line corresponds to $q_\mathrm{ballistic}=1$ with $t_\ast/\tau=15$ (25).
}\label{pppb2m_ball_a1024g_rin01_H1f_Q_Pa_rout100_alp05_cs5}\end{figure}

Since the magnetic field still increases exponentially as $e^{\gamma t}$, we have $B\sim e^{-\kappa r+\gamma t}=e^{-\kappa(r-c_\mathrm{front}t)}$,
where $c_\mathrm{front}=\gamma/\kappa=(\gamma\etat)^{1/2}$ is the front speed, as obtained from \Eq{Kappa}.
After an initial transient phase, the front radius begins to increase linearly with time; see \Fig{pppb2m_ball_a1024g_rin01_H1f_Q_Pa_rout100_alp05_cs5}.
This scaling is similar to that describing the propagation of fronts in epidemiological models \citep{Murray86}.
It also applies to the turbulent propagation of reactive fronts in combustion when the reaction speed exceeds the diffusion speed \citep{BHB11}.

\begin{figure*}\begin{center}
\includegraphics[width=\textwidth]{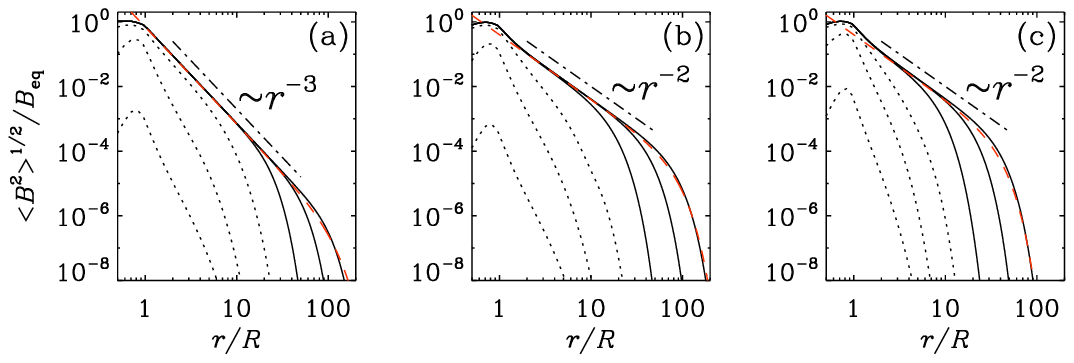}
\end{center}\caption{
Radial dependence of $\bra{\BB^2}^{1/2}$ at times $t/\tau_\mathrm{diff}=30$, 35, and 50
(dotted lines), and $t/\tau_\mathrm{diff}=100$, 300, and 1000 (solid lines), for
(a) a dipolar field with $C_\eta=50$ (Run~A),
(b) a quadrupolar field with $C_\eta=50$ (Run~B), and
(c) a quadrupolar field with $C_\eta=10$ (Run~C).
The asymptotic falloffs $\propto r^{-3}$ for the dipolar field
and $\propto r^{-2}$ for the quadrupolar fields are marked with dashed-dotted lines.
The red lines denote fits given by \Eq{eq:Bfit}, using the $\qdiff$ values listed in \Tab{TSummary}.
}\label{pprof_comp}\end{figure*}

\subsection{Magnetosphere radius}
\label{MagnetosphereRadius}

In \Fig{pprof_comp}, we plot the latitudinally averaged rms magnetic field for dipolar and quadrupolar boundary conditions.
For the quadrupolar case, we also include a model with a fivefold smaller magnetic diffusivity in the exterior.
These three models were also discussed in \cite{Ghosh+25}.
It turns out that, during the early kinematic growth phase, the magnetic
field grows and spreads in a similar manner, regardless of its hemispheric parity
and the value of the exterior magnetic diffusivity.
After the magnetic field has saturated in the dynamo-active region, it continues to spread
as $r^{-3}$ in the dipolar case, but as $r^{-2}$ in the two quadrupolar cases.
This behavior can clearly be seen in \Fig{pprof_comp}.
Note also that, during the kinematic phase, the fields decays more rapidly with radius than in the saturated phase.
Indeed, this represents a transient phase in which the magnetic field transitions from
an exponential radial decay, $\propto e^{-\kappa r}$, to a power-law decay, $\propto r^{-n}$,
within a shell around the dynamo, which subsequently becomes the magnetosphere.

The radial power-law decay extends from the end of the dynamo-active region at $r=R$ to a radius
$r_\ast(t)$ which we defined as the front radius in \Eq{FrontRadius}.
This radius grows diffusively with time after a reference time $t_\ast$.
As already demonstrated by \cite{Ghosh+25},
the sphere of radius $r_\ast(t)$ can be interpreted as a magnetosphere.
For the nondimensional coefficient $\qdiff$ in \Eq{FrontRadius},
the empirically determined value is close to 2---analogously to the law of Brownian diffusion \citep{Einstein05}.
In \Fig{pprof_comp}, we also compare the modulus of the latitudinally averaged magnetic field at the final time step with a fit of the form
\begin{equation}
\meanB(r,t)=B_0\,\left(\frac{r}{1\kpc}\right)^{-n} \exp\{-\textstyle{\frac{1}{2}}[r/r_\ast(t)]^2\},
\label{eq:Bfit}
\end{equation}
where $B_0$ characterizes the magnetic field strength inside the magnetosphere and $n$ is the radial decay exponent.
We find that the value of $\qdiff$ is around 4.9 for the dipolar configurations and around 2.5 for the quadrupolar configurations;
see \Tab{TSummary}.
This demonstrates that the radius of the magnetosphere scales with the exterior magnetic diffusivity, $\eta_{\rm turb}^{\rm ext}$,
and is also independent of the diffusivity contrast, $C_\eta$.
Even the case $C_\eta=1$ results in the same scaling that is expected for the value $\eta_{\rm turb}^{\rm ext}$,
which is then also equal to $\eta_{\rm turb}^{\rm ext}$ in that case.

Following \cite{Ghosh+25}, we determine the instantaneous front
radius $r_\ast(t)$ as a weighted integral:
\begin{equation}
\left.r_\ast(t)=\int r^{n+1} \meanB(r,t)\,d r\right/\!\int r^{n} \meanB(r,t)\,d r.
\label{eq:raverage}
\end{equation}
The diffusive $r_\ast(t)\propto t^{1/2}$ scaling is confirmed for all runs where $\etaT=\const=\eta_{\rm turb}^{\rm ext}$ in the exterior;
see Run~D (similar to Run~A), and Run~J (similar to Run~B).
The similarity between these pairs of runs also shows that neither the value of $C_\eta$ nor
the inclusion of feedback from the mean flow has a noticeable effect.
Since the latter would act in a manner analogous to the aforementioned magneto-frictional approach,
we conclude that the magnetic fields are already close to being force-free.

\subsection{Radially varying diffusivity}

\begin{figure}\begin{center}
\includegraphics[width=\columnwidth]{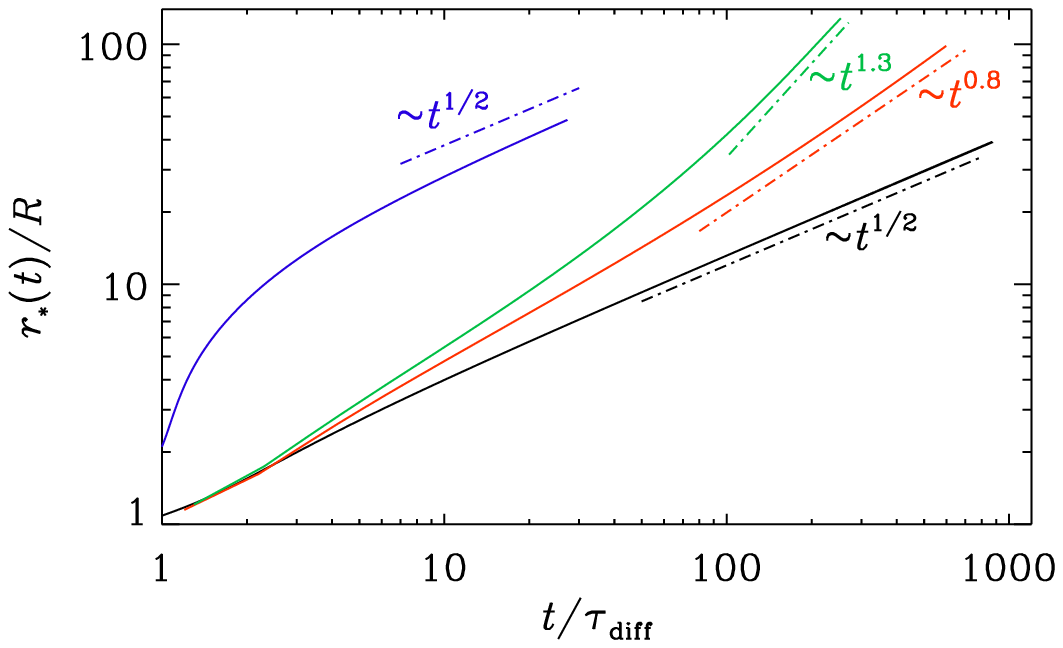}
\end{center}\caption{
Comparison of $r_\ast(t)$ for Runs~B (blue), G (black), G' (red), and G'' (green).
}\label{pppb2m_r_comp}\end{figure}

The local front speed depends on the value of $\etaT$.
When this is a function of $r$, the front speed can increase in regions where $\etaT$ is larger.
Thus, for a radially increasing profile, i.e., $m>0$ in \Eq{power_eta}, we
expect the front speed to increase with $r$.
This behavior is demonstrated in \Fig{pppb2m_r_comp}, where we compare $r_\ast(t)$ for Runs~G' and G'' with $m=1$ and 1.5 to the case $m=0$.
Here, we adopt $r_0/R=10$ in \Eq{power_eta}.
At large radii, we find a steeper temporal growth with $\propto t^{0.8}$ for $m=1$ and $\propto t^{1.3}$ for $m=1.5$.
However, the shallow inverse-quadratic radial decay of the mean magnetic field for the quadrupolar case remains approximately valid.
In \App{PowerLawGrowth}, we demonstrate that more precise measurements yield slightly steeper radial decays,
with $n\approx2.16$ and 2.23 for models with $m=1$ and 1.5, respectively.

The temporal scalings of $r_\ast(t)$ for different values of $m$ give an approximate idea about the sensitivity.
We see that the lines remain still somewhat curved, indicating the absence of a true power-law scaling.
If the turbulent magnetic diffusivity outside the galaxy were to be highly variable, we might expect a corrugated magnetosphere.
Conversely, if the magnetosphere is not strongly corrugated, this would suggest a more nearly constant distribution of $r_\ast$.
In either case, however, the spatial scaling of the magnetic field within the magnetosphere remains
approximately $\propto r^{-2}$ for the quadrupolar configuration.

\subsection{Comments on nonuniform diffusive scalings}

It is tempting to refer to a diffusive scaling as one that scales as $t^{1/2}$ with time.
Conversely, one might refer to deviations from this behavior as nondiffusive.
This is, however, misleading, because the expansion speed remains limited by the local magnetic diffusivity.
We can see this by inserting \Eq{power_eta} into \Eq{FrontRadius} for the saturated phase, and assuming $r\gg r_0$, which yields
\begin{equation}
r_\ast(t)/r_0\approx\left[q_\mathrm{diff} \, \etaT^\mathrm{ext} \, (t-t_\ast)/r_0^2\right]^{1/(2-m)}.
\end{equation}
Thus, if correct, we might asymptotically expect a temporal growth $\propto t$ for $m=1$ (instead of the numerically obtained $t^{0.8}$ behavior)
and $\propto t^2$ for $m=1.5$ (instead of $t^{1.3}$).
This discrepancy is consistent with the fact that the lines in \Fig{pppb2m_r_comp} remain somewhat curved,
indicating that a true powerlaw scaling has not yet been reached.

We emphasize that the accelerated expansion does not affect our conclusion that magnetic fields cannot significantly extend into cosmic voids.
This would require unrealistically large levels of turbulence in the voids.
We return to this point later in the paper when making a more quantitative comparison with the vacuum case.

\subsection{Radial power-law scaling within the magnetosphere}

To study the radial power-law behavior for dipolar and quadrupolar configurations
in more detail, we now express the latitudinal dependence in terms of Legendre polynomials.
An axisymmetric magnetic field can be written as \citep{Mof78, Par79, KR80}
\begin{equation}
\meanBB=\pphi \meanB_\phi+\nab\times\pphi \meanA_\phi,
\end{equation}
where $\meanA_\phi(r,\theta,t)$ and $\meanB_\phi(r,\theta,t)$ are independent of
each other outside the dynamo region and characterize the poloidal and toroidal magnetic field decay in the radial direction.
These functions can be expanded as\footnote{In general, axisymmetric and
nonaxisymmetric magnetic fields can be written as
\begin{equation}
\meanBB=\nab\times\rr T+\nab\times\nab\times\rr S,
\end{equation}
where $T(r,\theta,\phi,t)$ and $S(r,\theta,\phi,t)$ are the superpotentials
for the three-dimensional toroidal and poloidal fields, respectively.
In axisymmetry, this implies that $B_\phi=-\partial T/\partial\theta$, and
if $T(r,\theta)=P_\ell(\cos\theta)$, then $B_\phi=-P_\ell^1(\cos\theta)$.
Likewise, using $A_\phi=-\partial S/\partial\theta$, and if
$S(r,\theta)=P_\ell(\cos\theta)$, then $A_\phi=-P_\ell^1(\cos\theta)$.}
\begin{equation}
\meanA_\phi=\sum a_\ell(r)\,P_\ell^1(\cos\theta),\quad
\meanB_\phi=\sum b_\ell(r)\,P_\ell^1(\cos\theta).
\end{equation}
Dipolar (quadrupolar) magnetic fields have odd (even) values of $\ell$ in
$a_\ell(r)$ and even (odd) values of $\ell$ in $b_\ell(r)$.

\begin{figure*}\begin{center}
\includegraphics[width=.8\textwidth]{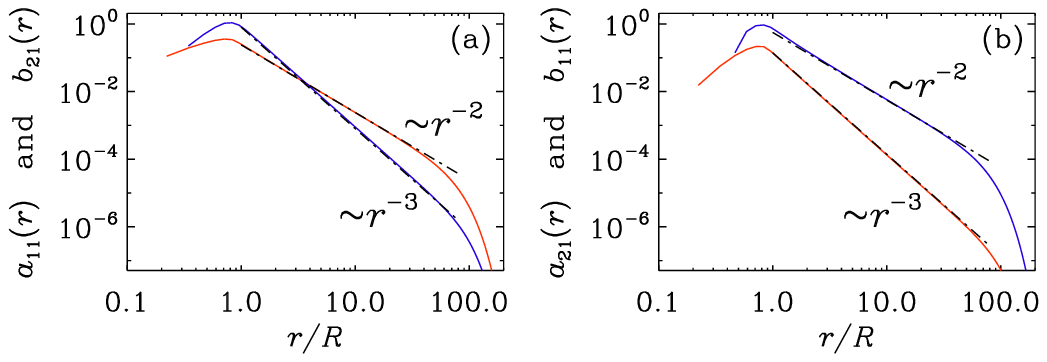}
\end{center}\caption{
Radial dependence of (a) $a_{11}(r)$ and $b_{21}(r)$ for the dipolar case (Run~A), and
(b) $a_{21}(r)$ and $b_{11}(r)$ for the quadrupolar case (Run~B).
The asymptotic slopes are $a_{11}/(\Beq R)\approx0.052\,(r/R)^{-2}$ and $b_{21}/\Beq\approx0.3\,(r/R)^{-3}$ for the dipole,
and $b_{11}/\Beq\approx0.132\,(r/R)^{-2}$ and $a_{21}/(\Beq R)\approx0.037\,(r/R)^{-3}$ for the quadrupole;
these are marked with dashed--dotted lines.
For a vacuum field, $b_{11}$ would be zero.
The red and blue lines indicate the scalings of $\meanA_\phi$ and $\meanB_\phi$, respectively, which are opposite for the dipolar and quadrupolar cases.
}\label{pradial_comp}\end{figure*}

We compute the coefficients $a_\ell(r)$ and $b_\ell(r)$ from our simulations as
\begin{equation}
a_\ell(r)=N_\ell\int \meanA_\phi(r,\theta) \,P_\ell^1(\cos\theta)\,\dd\cos\theta,
\end{equation}
\begin{equation}
b_\ell(r)=N_\ell\int \meanB_\phi(r,\theta) \,P_\ell^1(\cos\theta)\,\dd\cos\theta,
\end{equation}
where $N_\ell=(2\ell+1)\,(\ell-1)!/(\ell+1)!$ is a normalization factor.
In particular, $N_1=3/2$ and $N_2=5/6$.
In \Fig{pradial_comp}, we plot the radial dependence of $a_{11}(r)$ and $b_{21}(r)$ for the dipole case
and of $a_{21}(r)$ and $b_{11}(r)$ for the quadrupolar case.
We clearly see the asymptotic slopes $a_{11}\propto r^{-2}$ and $b_{21}\propto r^{-3}$ for the dipole,
and $b_{11}\propto r^{-2}$ and $a_{21}\propto r^{-3}$ for the quadrupole.

Note that for the dipolar and quadrupolar solutions, the roles of $A_\phi$ and $B_\phi$ are interchanged; see \Fig{pradial_comp}.
In particular, the $r^{-2}$ behavior for the $\meanA_\phi$ of the dipole is now seen for the $\meanB_\phi$ field of the quadrupole,
while the $r^{-3}$ behavior for the $\meanB_\phi$ of the dipole is now seen for the $\meanA_\phi$ field of the quadrupole.
Thus, although the poloidal component of a quadrupolar configuration still decays as $r^{-4}$,
its toroidal component $\meanB_\phi$ decays as $\meanA_\phi$ for the dipolar configuration, i.e., as $r^{-2}$.

In the following, we discuss the individual components of the magnetic field in more detail.
The two components of the poloidal field are given by $B_r=\DDt A_\phi$ and $B_\theta=-\DDr A_\phi$,
where $\DDt=r^{-1}\sin^{-1}\!\theta\,\partial_\theta(\sin\theta\,\cdot)$, and $\DDr=r^{-1}\partial_r(r\,\cdot)$
are differential operators.

\begin{figure*}\begin{center}
\includegraphics[width=.8\textwidth]{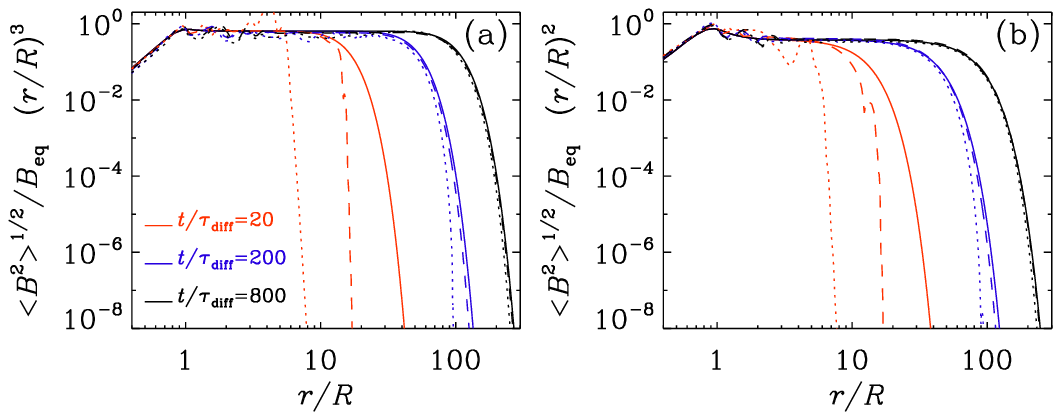}
\end{center}\caption{
Radial magnetic field profiles compensated (a) by $r^3$ for the dipolar case (Runs~D--F), and (b) by $r^2$ for the quadrupolar case (Run~G--I).
For the solid lines, the displacement current is neglected (Runs~D and G), while for the dashed and dotted lines it is included
with $c R/\etaT=1$ (Runs~E and H) and 0.5 (Runs~F and I), respectively.
The red, blue, and black lines correspond to the times $t/\tau_\mathrm{diff}=20$, 200, and 800.
}\label{pprof_comp_ee}\end{figure*}

\subsection{Lowest-order dipole}
\label{LowestOrderDipole}

Employing our lowest-order dipole approximation, we find the two fields
to be proportional to
\begin{equation}
\meanA_\phi=a_\mathrm{D}\,r^{-2}\,P_1^1(\cos\theta),\quad
\meanB_\phi=b_\mathrm{D}\,r^{-3}\,P_2^1(\cos\theta).
\end{equation}
Therefore, since $P_1^1(\cos\theta)=-\sin\theta$ and $P_1(\cos\theta)=\cos\theta$, we have
\begin{equation}
\meanB_r=-2a_\mathrm{D}\,r^{-3}\,\cos\theta\equiv-2a_\mathrm{D}\,r^{-3}\,P_1(\cos\theta),
\end{equation}
\begin{equation}
\meanB_\theta=-a_\mathrm{D}\,r^{-3}\,\sin\theta\equiv a_\mathrm{D}\,r^{-3}\,P_1^1(\cos\theta).
\end{equation}
For our model, we find $a_\mathrm{D}\approx0.20\,\Beq R^3$ and $b_\mathrm{D}\approx0.60\,\Beq R^3$.

\subsection{Lowest-order quadrupole}
\label{LowestOrderQuadrupole}

Our lowest-order quadrupole is proportional to
\begin{equation}
\meanA_\phi=a_\mathrm{Q}\,r^{-3}\,P_2^1(\cos\theta),\quad
\meanB_\phi=b_\mathrm{Q}\,r^{-2}\,P_1^1(\cos\theta).
\end{equation}
Since $P_2^1(\cos\theta)=-3\sin\theta\cos\theta$ and $P_2(\cos\theta)=(3\cos^2\theta-1)/2$, we have
\begin{equation}
\meanB_r=3a_\mathrm{Q}\,r^{-4}\,(1-3\cos^2\theta)\equiv-6a_\mathrm{Q}\,r^{-4}\,P_2(\cos\theta),
\end{equation}
\begin{equation}
\meanB_\theta=-6a_\mathrm{Q}\,r^{-4}\,\sin\theta\cos\theta\equiv 2a_\mathrm{Q}\,r^{-4}\,P_2^1(\cos\theta).
\end{equation}
For our model, we find $a_\mathrm{Q}\approx0.09\,\Beq R^4$ and $b_\mathrm{Q}\approx0.49\,\Beq R^2$.

\subsection{Current density profiles}
\label{CurrentDensityProfiles}

Next, we compute $\mu_0\meanJJ=\nab\times\meanBB$, which is given by
\begin{equation}
\mu_0\meanJJ=\pphi \mu_0\meanJ_\phi+\nab\times\pphi \meanB_\phi,
\end{equation}
where
\begin{equation}
\mu_0\meanJ_\phi=-(\DD_r^2+{\cal D}_\theta^2) \meanA_\phi,
\end{equation}
is proportional to the toroidal current density.
Again, $\Drr\equiv\DDr\DDr$ and $\dDt\equiv\ddt\DDt$ where
$\ddt=r^{-1}\partial_\theta$ is a convenient shorthand.

For the lowest-order dipole with $A_\phi=a_\mathrm{D}\,r^{-2}\,P_1^1(\cos\theta)$, we have
\begin{equation}
\DD_r^2 (-r^{-2}\sin\theta)=-2\,r^{-4}\sin\theta
\end{equation}
and
\begin{equation}
{\cal D}_\theta^2 (-r^{-2}\sin\theta)=2\,r^{-4}\sin\theta,
\end{equation}
and therefore $\meanJ_\phi=0$.
Thus, we have $\mu_0\meanJJ=\nab\times\pphi \meanB_\phi$, and therefore
the Lorentz force $\meanJJ\times\meanBB$ is given by
\begin{equation}
\meanJJ\times\meanBB=(\nab\times\pphi \meanB_\phi)\times(\pphi \meanB_\phi+\nab\times\pphi \meanA_\phi)/\mu_0,
\end{equation}
and has only a purely toroidal component whose divergence vanishes.
Therefore, such a field is nearly force-free, except for the $\phi$ component,
which drives only rotational motion.
This agrees with our conclusion at the end of \Sec{MagnetosphereRadius}.

For the quadrupole with $\meanA_\phi\propto r^{-3}\sin\theta\cos\theta$, we have
\begin{equation}
\DD_r^2 r^{-3}=6\,r^{-5}
\end{equation}
and
\begin{equation}
{\cal D}_\theta^2 \sin\theta\cos\theta
=-6\,r^{-2}\sin\theta\cos\theta.
\end{equation}
Therefore, again $\meanJ_\phi=0$.

For the quadrupole, we have
\begin{equation}
D^2\meanB_\phi=0,
\end{equation}
in a manner similar to $\meanA_\phi$ for the dipolar field.
However, unlike for $\meanA_\phi$, where $D^2\meanB_\phi=\mu_0\meanJ_\phi$
is the toroidal current density, in the case of a quadrupolar field
$D^2\meanB_\phi$ has no physically obvious meaning.
What is more relevant in a conducting medium is the Lorentz force,
$\meanJJ\times\meanBB$, which has both poloidal and toroidal components.
Its toroidal component, arising from the poloidal components
$\meanJJ_\mathrm{pol}$ and $\meanBB_\mathrm{pol}$, is nonvanishing,
but it only drives azimuthal differential rotation
and becomes extremely weak due to its rapid radial decay $\propto r^{-5}$.
The poloidal Lorentz force has the components
$\meanJJ_\mathrm{pol}\times\meanB_\phi\pphi$ and
$\pphi D^2\meanA_\phi\times\meanBB_\mathrm{pol}/\mu_0$.
Using $\mu_0\meanJJ_\mathrm{pol}=\nab\times\meanB_\phi\pphi$ and $\meanBB_\mathrm{pol}=\nab\times\meanA_\phi\pphi$, we have
$(\nab\times\meanB_\phi\pphi)\times\meanB_\phi\pphi=-\varpi^{-1}\nab(\varpi\meanB_\phi^2)$,
which yields a radial force that also decays rapidly as $r^{-5}$.
Thus, except for a term $\meanB_\phi^2\nab\ln\varpi^2$, the Lorentz force can in principle be balanced by a pressure gradient.
The residual $\meanB_\phi^2\nab\ln\varpi^2$ term, on the other hand, can be balanced by the centrifugal force.
In this sense, the quadrupolar configuration is nearly force-free.

As already stated above, outside the dynamo region $\meanA_\phi$ and $\meanB_\phi$ act independently of each other,
and each component is subject to diffusion.
Thus, unlike in the dynamo interior, where $\meanB_\phi$ is coupled to $\meanA_\phi$ through induction effects such as differential rotation,
$\meanB_\phi$ here expands purely through (turbulent) magnetic diffusion and can therefore adopt the same behavior as $\meanA_\phi$.
In the present case, this means that for the quadrupolar configuration, $\meanB_\phi\propto r^{-2}$, 
just like the $\meanA_\phi\propto r^{-2}$ behavior for the dipolar configuration.

\subsection{Effects of the displacement current}

As we have seen from the Cartesian simulations in \App{DispersionRelation},
the inclusion of the displacement current can lead to a reduced growth rate $\gamma$; see \App{DispersionRelation}.
It also can only decrease the front speed, $c_\mathrm{front}=\gamma/\kappa$.
As a consequence, the front speed always remains below the speed of light.

\begin{figure*}\begin{center}
\includegraphics[width=.8\textwidth]{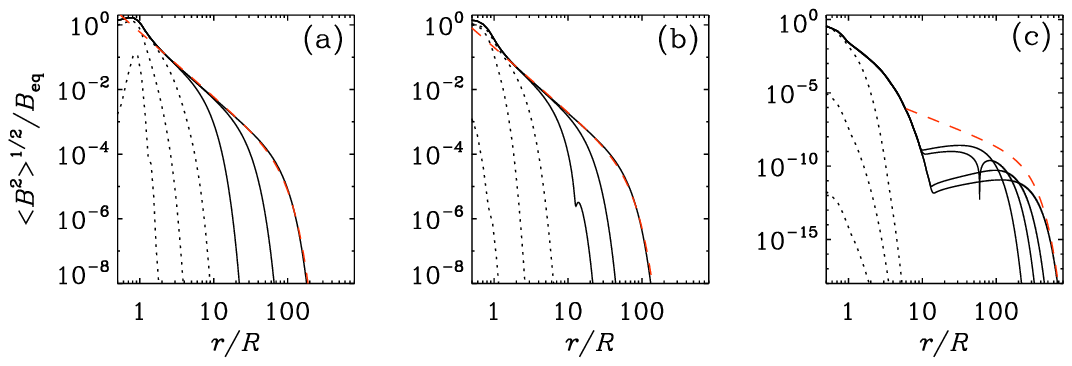}
\end{center}\caption{
Radial magnetic field profiles for Runs~J, K, and L.
For Run~J, we have $r_\mathrm{in}/R=0.2$ instead of 0.1, while 
for Runs~K and L, we have $h/R=0.5$ and 0.2, respectively.
The times are (a) $t/\tau_\mathrm{diff}=0.1$, 0.3, and 1.5 for the dotted lines, and 10, 100, and 942 for the solid lines;
(b) 0.1, 0.2, and 0.8 for the dotted lines, and 10, 50, and 530 for the solid lines;
and (c) 0.2, 0.4, and 0.6 for the dotted lines, and 500, 1200, 2700, and 5990 for the solid lines.
}\label{pprof_comp2}\end{figure*}

As alluded to in the introduction, the values of the magnetic diffusivity would need to be extremely large if one wanted to emulate the conditions of a vacuum.
In vacuum, an electromagnetic wave propagates at the speed $c=300,000\kms$, which implies a propagation distance
$L=c t_\mathrm{H}\approx5\Gpc$ in one Hubble time, $t_\mathrm{H}\approx14\Gyr$.
This corresponds to a reference diffusivity of about $\eta_\mathrm{ref}\equiv L^2/t_\mathrm{H}=c^2 t_\mathrm{H}\approx2\times10^{12}\kpc\kms$,
which is nearly $10^9$ times larger than the value $\etat=3000\kpc\kms$ used by \cite{Ghosh+25}.
Such large values are unrealistic.\footnote{
Even if the turbulent velocity was equal to the speed of light, it
would require a correlation length of more than $70\Gpc$ to reach
a turbulent diffusivity of $10^{12}\kpc\kms$.
}

Nevertheless, it is of interest to see how the spherical expansion of the magnetic field is affected by the finite speed of light
in a medium with such an extremely large effective magnetic diffusivity.
In the following, we consider the values $cR/\eta_\mathrm{ref}=1$ and 0.5.
Again, these ratios are chosen to be unrealistically small.
This is done just to see the qualitative and quantitative effects of including the Faraday displacement current.
We again choose $C_\eta=1$, which, as demonstrated above, does not affect the radial expansion behavior compared to cases with larger values of $C_\eta$.
The results for the spherical models are shown in \Fig{pprof_comp_ee}.
For these models, we have neglected the evolution of the mean flow, i.e., we solve only \Eqs{dAdt}{dE2dt}.
We find that only at early times is the solution is noticeably affected by the finite speed of light---even for $c R/\etaT=0.5$.
In these cases, as explained in \Sec{Details}, we characterize the front radius by the radius $r_{0.01}$,
where the compensated magnetic field has dropped to 0.01 times its value inside the magnetosphere.
From \Tab{TSummary}, we see that $r_{0.01}/ct<1$ is satisfied in all cases with finite values of $cR/\etaT$.
This ratio varies between 0.8 and 0.6 for our dipolar configurations (Runs~E and F)
and between 0.7 and 0.6 (Runs~H and I) for our quadrupolar configurations.
Thus, $r_{0.01}$ remains close to, but always below, the Hubble radius.

\subsection{Effects of changing $r_\mathrm{in}$ and $h$}

In \Fig{pprof_comp2}, we examine the effects of varying the ratios $r_\mathrm{in}/R$ and $h/R$ (Runs~J--L).
To ensure that the dynamo remains excited despite its smaller volume, we have here also increased the value of $C_\alpha$ from 25 to 50.
We find that the radial falloff is basically unchanged when increasing the value of $r_\mathrm{in}/R$ from 0.1 (Run~G) to 0.2 (Run~J).
By contrast, when decreasing the value of $h/R$ from 1 (Run~G) to $0.5$ (Run~K) and $0.2$ (Run~L) leads to more complex radial profiles,
including the development of radial oscillations.
In addition, there is even an indication of a reduction in the magnetic field strength in the wake of the front.
To examine this more clearly, we run the case with $h/R=0.2$ for a much longer time.
The results suggest that the magnetic field forms an azimuthal ring, or at least a shell, rather than a filled sphere.
This complication obscures the otherwise simple radial power-law behavior discussed above,
and also motivates our focus on the case $H/R=1$.
It is important to note, however, that the position of the exponential falloff at the edge of the magnetosphere remains unchanged,
as evidenced by the nearly constant value of the parameter $\qdiff$ listed in \Tab{TSummary} and
by the fits presented in \Fig{pprof_comp2} for the final time step.

\section{Observational implications}
\label{Observational}

To compare with future observations, \cite{Ghosh+25} computed the rotation measure (RM).
Assuming a radial power-law dependence for the thermal electron density, $n_\mathrm{th}\propto r^{-s_\mathrm{th}}$,
they found that RM decreases with radius as $r^{-2-s_\mathrm{th}}$ for the dipolar solution
and as $r^{-1-s_\mathrm{th}}$ for the quadrupolar solution.
Measuring such radial profiles can therefore help to test our predicted scalings.

Another observational diagnostic is the resulting synchrotron emission.
The synchrotron intensity $I(x,z,\lambda)$ and the complex polarization
$\mathcal{P}(x,z,\lambda)\equiv Q+\ii U$ are obtained by line-of-sight integration.
Here, $Q$ and $U$ are the Stokes parameters characterizing linear polarization, and $\lambda$ is the wavelength.

\begin{figure*}[t!]\begin{center}
\includegraphics[width=\textwidth]{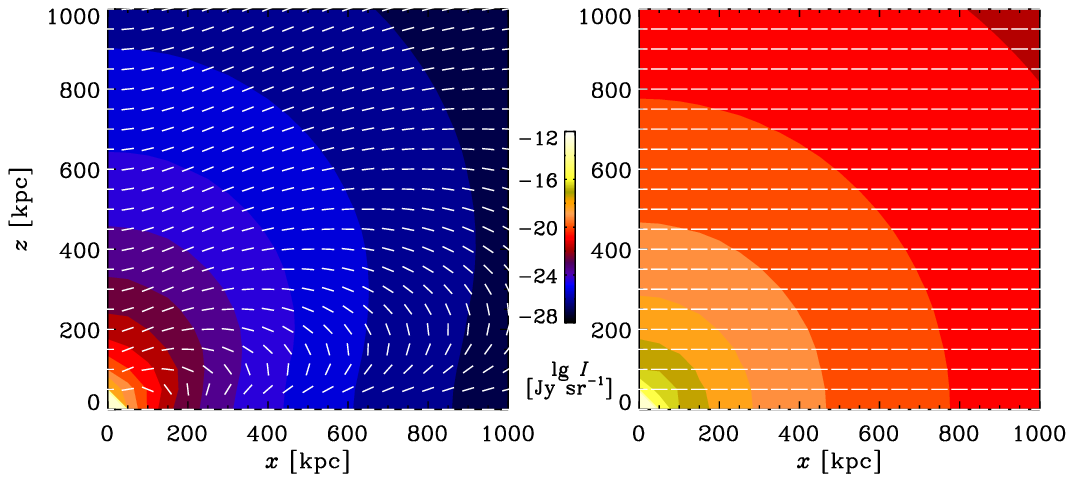}
\end{center}\caption{
Polarization vectors $(Q,U)$ superimposed on a color-scale representation of the logarithmic intensity $I(x,z)$ (in Jy/sr)
for Runs~A (dipolar configuration, left) and B (quadrupolar configuration, right).
}\label{ppol-int-sideways}\end{figure*}

To illustrate the basic properties of $I$, $Q$, and $U$, 
we adapt the expression for the radial scaling of the magnetic field given by \Eq{eq:Bfit} to the
lowest-order dipolar and quadrupolar configurations discussed in \Secs{LowestOrderDipole}{LowestOrderQuadrupole}.
We then compute the components
\begin{equation}
\meanB_x=\sin\theta\cos\phi \,\meanB_r+\cos\theta\cos\phi \,\meanB_\theta-\sin\phi \,\meanB_\phi,
\label{eq:Bx}
\end{equation}
\begin{equation}
\meanB_z=\cos\theta \,\meanB_r-\sin\theta \,\meanB_\theta,
\label{eq:Bz}
\end{equation}
in Cartesian coordinates, where $\phi$ is the azimuthal angle and $\theta$ is the colatitude.
The synchrotron intensity $I(x,z)$ projected onto the $x$--$z$ plane is given by
\begin{equation}
I(x,z,\lambda)=\int_{-L}^L \epsilon(x,y',z,\lambda)\,\dd y',
\end{equation}
where $\epsilon\propto n_\mathrm{CR} \meanB_\perp^{(\ggamma+1)/2} \lambda^{(\ggamma-1)/2}$ is the emissivity,
$n_\mathrm{CR}$ is the cosmic-ray density, $\lambda$ is the wavelength, and $\ggamma$ is the spectral index of the momentum distribution \citep{GS65}.
Realistic values for $\ggamma$ lie in the range 2.8--3.2, as a result of the combination of the continuous injection
of secondary electrons (by hadronic collisions), the injection of primary electrons by supernova-driven shocks,
and the modulation by energy-dependent diffusion \citep[e.g.,][]{Beck91, 2021MNRAS.505.3273W,2025ApJ...989..140A}.
Here, for simplicity, we use $\ggamma=3$, in which case $\epsilon\propto \meanB_\perp^2$, and the complex intrinsic polarization
is simply proportional to $\mathcal{B}_\perp^2$, where $\mathcal{B}_\perp=\meanB_x+\ii\meanB_z$ is the
complex magnetic field in the plane of the sky; see \cite{BS14} for earlier work utilizing this formalism.

We neglect here the effects of Faraday rotation, which would introduce a factor
$\exp(2\ii\phi\lambda^2)$ under the integral of $\mathcal{P}$.
This is justified by the results of \citet{Ghosh+25}, who showed that Faraday rotation is negligible outside the galactic core.
Here, $\phi(x,y,z)=-K\int_0^y (n_\mathrm{th}\meanB_y)(x,y',z)\,\dd y'$ is the Faraday depth and $K$ is a known constant.
Neglecting Faraday rotation corresponds to taking the limit $\lambda\to0$.

In \Fig{ppol-int-sideways}, we present synthetic maps of the synchrotron intensity $I(x,z)$
for the same dipolar and quadrupolar configurations as in \cite{Ghosh+25}, corresponding to Runs~A and B, respectively.
For $n_\mathrm{CR}$, we assume for simplicity a steady radial profile similar to $n_\mathrm{th}$, which we write here as
$n_\mathrm{CR}\propto r^{-s_\mathrm{CR}}$, analogous to the expression used for $n_\mathrm{th}$ used by \cite{Ghosh+25}.
Here, we adopt $s_\mathrm{CR}=1$, although it should be noted that radiative losses may complicate this behavior
in reality \citep[e.g.,][]{2021ApJ...907L..11P,2022MNRAS.515.4229P,2025ApJ...989..140A}.
To obtain a quantitative estimate of the magnetospheric emission, we use
\citep[][see also \App{DerivationEq39}]{Rybicki+Lightman86}
\begin{equation}
\epsilon\,\dd y \approx\frac{5\,\mu\mathrm{Jy}}{\mathrm{sr}}
\left(\frac{n_\mathrm{CR}^{(100\kpc)}}{10^{-7}\cm^{-3}}\right)
\left(\frac{\Beq}{1\uG}\right)^2
\left(\frac{\lambda}{2\m}\right)
\left(\frac{\dd y}{1\kpc}\right),
\label{Eq39}
\end{equation}
where $n_\mathrm{CR}^{(100\kpc)}$ is the cosmic-ray density at a galactocentric distance of $100\kpc$,
assuming that $s_\mathrm{CR}=1$ remains valid at that distance, which in turn depends on the long-term integrated dispersal of cosmic rays around galaxies and within large-scale structures \citep[see, e.g.,][for recent numerical studies]{Vazza+25}.

In \Fig{ppol-int-sideways}, we overlay the polarization vectors $(Q,U)$.
For a power-law dependence of the cosmic-ray number density, $n_\mathrm{CR}\propto r^{-s_\mathrm{CR}}$,
assuming $\meanB_\perp\propto r^{-n}$ with $n=2$ for quadrupoles and $n=3$ for dipoles,
the intrinsic emissivity becomes proportional to $n_\mathrm{CR}\meanB_\perp^2\propto r^{-s_\mathrm{CR}-2n}$; see \Eq{eq:Bfit}.
Line-of-sight integration adds another power, yielding $r^{-s_\mathrm{CR}-2n+1}$.
Thus, for quadrupolar (dipolar) configurations, we find $I\propto r^{-3-s_\mathrm{CR}}$ ($\propto r^{-5-s_\mathrm{CR}}$).

We find that the polarization vectors are predominantly in the $x$-direction.
This is a commonly observed feature related to the dominance of the toroidal magnetic field; see, e.g., \cite{BDMSST93}
and \cite{Elstner+95} for earlier work where, unlike the present case, a radial wind was included.
Only in the dipolar configuration does the toroidal field change sign at the midplane, leading to
clear deviations from the otherwise nearly perfect alignment of the polarization vectors with the $x$ direction.

As expected, the radial falloff of the synchrotron intensity is still rather steep---even for the quadrupolar configuration.
Let us therefore now discuss whether our predicted intensity distributions are observable in real galaxies.
For example, an $8$ hour integration with the LOFAR High Band Antenna (HBA) at $\approx150\MHz$ typically reaches a noise level of
$3 \sigma_{\rm rms} \approx 1 \uJy/\rm arcsec^2=2.35 \cdot 10^{-17} \Jy/\sr$.
Based on \Fig{ppol-int-sideways}, a fraction of the predicted radio emission should be detectable in both cases by radio telescopes.
However, for the dipolar configuration, the detectable region is likely confined within a $\leq 200 \rm ~kpc$ radius,
whereas for the quadrupolar case a much larger region may be detectable.
A quantitative prediction requires fixing the radial distribution of cosmic rays, which can currently only roughly be guessed at this.
In real galaxies, this distribution is likely affected by the presence of large-scale outflows,
especially at larger radii where the inertia of cosmic rays is expected to become dynamically dominant \citep[e.g.,][]{2024ApJ...976..136C}.
Interestingly, however, the differences in the orientation of the polarization vectors is relatively insensitive to
the distribution of relativistic electrons.
This is unfortunate, because the orientation of the polarization vectors could provide a more direct probe of the large-scale distribution of magnetic fields.
Intriguingly, preliminary work using LOFAR observations may already have yielded results that are possibly consistent with our prediction of a finite magnetosphere \citep{Oei}.

Distinguishing whether the magnetosphere is produced by a galactic wind or by turbulent magnetic diffusion is not straightforward,
as both mechanisms lead to the same $r^{-2}$ flux-conserving radial decay.
However, combined spectral and polarimetric observations at large radii have been shown to provide a sensitive diagnostic
of cosmic-ray (re)acceleration processes \citep[e.g.,][]{2025A&A...696A.226T}.
Such cosmic rays are expected in the presence of outflows \citep[e.g.,][]{2024ApJ...976..136C}, but not necessarily in the case of the extended magnetosphere studied in this work.

\section{Conclusions}
\label{Conclusions}

We have studied the radial spreading of a dynamo-generated magnetic
field into its electrically conducting surroundings.
Unlike in conventional dynamo theory, where the exterior field is assumed
to be a current-free potential field, our solution contains electric currents and is
only approximately force-free.
As already reported by \cite{Ghosh+25}, the quadrupolar field has a toroidal component
that decays with radius as $r^{-2}$, which is slower than the
$r^{-3}$ decay of the dipolar poloidal component in the far field of the dynamo.
To the best of our knowledge, such a slow radial magnetic field decay has not been reported previously.

An important difference from earlier work on force-free magnetic fields in dynamo exteriors is that we study the time-dependent case using
direct numerical simulations, in which a nearly force-free field builds up gradually in an electrically conducting medium.
As far as the formulation of the effective boundary conditions of the dynamo is concerned,
which was the main purpose of considering force-free exteriors \citep{Bonanno16, Bonanno+DelSordo17},
the finite time required to establish the exterior field may not be critical on such short length scales.
However, the slower radial decay and the sharp exponential falloff
marking the edge of the magnetosphere may be of observational interest \citep{Ghosh+25} and
could be tested with the new generation of radio observatories \citep[e.g.,][]{2022MNRAS.515.1158G}.
This feature may help distinguish such fields from dipolar configurations
on the one hand and from wind-governed ones on the other.
Galactic winds have indeed frequently been studied in simulations \citep{Aramburo2021, Aramburo2022, bondarenko2022account},
including cases in which a dynamo is coupled either to an imposed wind \citep{BDMSST93} or to a dynamically sustained one \citep{Perri+21, JB21}.

In the work of \cite{BDMSST93}, maps of synchrotron emission were presented,
but the displayed results do not show a clear determination of whether the radial falloff is characterized by a power law.
\cite{JB21} did not compute synchrotron emission, but they presented their results for $\meanB_\phi$
compensated by $r^2$.
This shows that, although their field configuration was dipolar, the presence of a wind alone can also
lead to a slow radial decay of $\meanB_\phi$.

In the present cases, an important clue to understanding the origin of the $r^{-2}$ decay of $\meanB_\phi$ 
comes from the correspondence between the $(\meanB_\phi,\meanA_\phi)$ pair for a quadrupolar configuration
and the $(\meanA_\phi,\meanB_\phi)$ pair for a dipolar configuration.
In both cases, the residual components decay as
\begin{equation}
\left.
\begin{array}{l}
(\meanB_\phi,\meanA_\phi)\cr
(\meanA_\phi,\meanB_\phi)
\end{array}
\right\}=
\left[r^{-2}\,\,P_1^1(\cos\theta),\; r^{-3}\,\,P_2^1(\cos\theta)\right],
\end{equation}
and their evolution is simply the result of diffusion.
Unlike the action of $\alpha$- and $\Omega$-effects in the dynamo interior, diffusion does not couple $\meanA_\phi$ and $\meanB_\phi$,
which make this correspondence is possible.

While these results are also borne out in cases where the displacement current is included,
we have shown that it can safely be ignored in all cases of interest.
Its main role lies in limiting the expansion speed in poorly conducting media rather than facilitating a radial expansion.

It is also interesting to note that the far-field magnetic field configurations are nearly force-free, in the sense that the resulting Lorentz force
can be balanced by pressure-gradient and centrifugal forces.
However, this aspect is likely subdominant, as the Lorentz force decays rapidly with radius and our results are essentially
independent of whether the momentum equation is included.

In the present work, we have made a number of simplifying assumptions that have helped produce clean results.
In particular, assuming a spherical rather than an oblate dynamo geometry have helps
produce a power-law behavior in the radial decay of the magnetic field.
It is possible that, at large distances from the dynamo region, this assumption becomes less critical.
However, the present results for oblate dynamo geometries rather suggest that the magnetosphere may instead take the form of a ring or a spherical shell.
The inclusion of differential rotation in the dynamo region would further complicate the results.
Preliminary studies have shown that the $r^{-2}$ falloff for the quadrupolar configuration may become steeper.
It remains unclear, however, how this changes at larger distances.
Further modifications are expected if a radial wind is allowed to develop.
In that case, the results may depend on the level of galactic activity driving such a wind.
Nevertheless, a slow radial falloff for quadrupolar fields may still survive under certain more realistic conditions.
As explained above, the resulting synchrotron emission may be of observational interest.
However, as already stressed by \cite{Ghosh+25}, the radii of galactic magnetospheres are not affected by this
and remain prohibitively small in view of scenarios such as those proposed by \cite{Garg+25}, in which the lower limits of void magnetization
are explained by the superposition of dipoles from the far field of galaxies.

\begin{acknowledgments}
We thank the referee for a constructive review.
We acknowledge stimulating discussions with Ruth Durrer, Deepen Garg, and Jennifer Schober.
This research was supported in part by the European Research Council through the ERC Synergy Grant COSMOMAG under grant No.\ 101224803,
the Swedish Research Council (Vetenskapsr{\aa}det) under grant No.\ 2025-05957,
the National Science Foundation under grant Nos.\ NSF AST-2307698, AST-2408411, and NASA Award 80NSSC22K0825. OG acknowledges support from a Wallenberg Academy Fellowship (PI Azadeh Fattahi). Additionally, OG is supported by the Swedish Research Council (Vetenskapsr{\aa}det) under contracts 2022-04283 and 2019-02337 and additionally by the G\"oran Gustafsson Foundation for Research in Natural Sciences and Medicine.  
AN is partially supported by the French National Research Agency (ANR) grant ANR-24-CE31-4686.
FV has been partially supported by Fondazione Cariplo and Fondazione CDP, through grant n$^\circ$ Rif: 2022-2088 CUP J33C22004310003 for the ``BREAKTHRU'' project.
We acknowledge the allocation of computing resources provided by the Swedish National Allocations Committee at the Center for
Parallel Computers at the Royal Institute of Technology in Stockholm.

\vspace{2mm}\noindent
{\em Software and Data Availability.}
The source code used for the simulations of this study,
the {\sc Pencil Code} \citep{PC}, is freely available on
\url{https://github.com/pencil-code}.
The simulation setups and corresponding input
and reduced output data are freely available on
\dataset[http://doi.org/10.5281/zenodo.18564321]{http://doi.org/10.5281/zenodo.18564321}.
\end{acknowledgments}

\bibliographystyle{aasjournal}
\bibliography{ref}

\begin{thebibliography}{}
\expandafter\ifx\csname natexlab\endcsname\relax\def\natexlab#1{#1}\fi
\providecommand{\url}[1]{\href{#1}{#1}}
\providecommand{\dodoi}[1]{doi:~\href{http://doi.org/#1}{\nolinkurl{#1}}}
\providecommand{\doeprint}[1]{\href{http://ascl.net/#1}{\nolinkurl{http://ascl.net/#1}}}
\providecommand{\doarXiv}[1]{\href{https://arxiv.org/abs/#1}{\nolinkurl{https://arxiv.org/abs/#1}}}

\bibitem[{Aramburo-García {et~al.}(2021)}]{Aramburo2021}
Aramburo-García, A., {et~al.} 2021, Mon. Not. R. Astron. Soc., 502, 6012,
  \dodoi{10.1093/mnras/stab391}

\bibitem[{Aramburo-García {et~al.}(2022)}]{Aramburo2022}
---. 2022, Mon. Not. R. Astron. Soc., 514, 2656, \dodoi{10.1093/mnras/stac1460}

\bibitem[{{Armillotta} {et~al.}(2025){Armillotta}, {Ostriker}, \&
  {Linzer}}]{2025ApJ...989..140A}
{Armillotta}, L., {Ostriker}, E.~C., \& {Linzer}, N.~B. 2025, \apj, 989, 140,
  \dodoi{10.3847/1538-4357/adea68}

\bibitem[{{Beck}(1991)}]{Beck91}
{Beck}, R. 1991, \aap, 251, 15

\bibitem[{{Bonanno}(2016)}]{Bonanno16}
{Bonanno}, A. 2016, \apjl, 833, L22, \dodoi{10.3847/2041-8213/833/2/L22}

\bibitem[{{Bonanno} \& {Del Sordo}(2017)}]{Bonanno+DelSordo17}
{Bonanno}, A., \& {Del Sordo}, F. 2017, \aap, 605, A33,
  \dodoi{10.1051/0004-6361/201731330}

\bibitem[{Bondarenko {et~al.}(2022)Bondarenko, Boyarsky, Korochkin, Neronov,
  Semikoz, \& Sokolenko}]{bondarenko2022account}
Bondarenko, K., Boyarsky, A., Korochkin, A., {et~al.} 2022, A\&A, 660, A80

\bibitem[{{Brandenburg} {et~al.}(1993){Brandenburg}, {Donner}, {Moss},
  {Shukurov}, {Sokoloff}, \& {Tuominen}}]{BDMSST93}
{Brandenburg}, A., {Donner}, K.~J., {Moss}, D., {et~al.} 1993, \aap, 271, 36

\bibitem[{{Brandenburg} {et~al.}(2011){Brandenburg}, {Haugen}, \&
  {Babkovskaia}}]{BHB11}
{Brandenburg}, A., {Haugen}, N. E.~L., \& {Babkovskaia}, N. 2011, \pre, 83,
  016304, \dodoi{10.1103/PhysRevE.83.016304}

\bibitem[{{Brandenburg} {et~al.}(1992){Brandenburg}, {Moss}, \&
  {Tuominen}}]{Bra+92}
{Brandenburg}, A., {Moss}, D., \& {Tuominen}, I. 1992, \aap, 265, 328

\bibitem[{{Brandenburg} \& {Stepanov}(2014)}]{BS14}
{Brandenburg}, A., \& {Stepanov}, R. 2014, \apj, 786, 91,
  \dodoi{10.1088/0004-637X/786/2/91}

\bibitem[{{Brandenburg} {et~al.}(1990){Brandenburg}, {Tuominen}, \&
  {Krause}}]{BTK90}
{Brandenburg}, A., {Tuominen}, I., \& {Krause}, F. 1990, Geophys. Astrophys.
  Fluid Dynam., 50, 95, \dodoi{10.1080/03091929008219875}

\bibitem[{{Chandrasekhar} \& {Kendall}(1957)}]{CK57}
{Chandrasekhar}, S., \& {Kendall}, P.~C. 1957, \apj, 126, 457,
  \dodoi{10.1086/146413}

\bibitem[{{Chiu} {et~al.}(2024){Chiu}, {Ruszkowski}, {Thomas}, {Werhahn}, \&
  {Pfrommer}}]{2024ApJ...976..136C}
{Chiu}, H.-H.~S., {Ruszkowski}, M., {Thomas}, T., {Werhahn}, M., \& {Pfrommer},
  C. 2024, \apj, 976, 136, \dodoi{10.3847/1538-4357/ad84e9}

\bibitem[{{Einstein}(1905)}]{Einstein05}
{Einstein}, A. 1905, Ann. Phys., 322, 549, \dodoi{10.1002/andp.19053220806}

\bibitem[{{Elstner} {et~al.}(1995){Elstner}, {Golla}, {R\"udiger}, \&
  {Wielebinski}}]{Elstner+95}
{Elstner}, D., {Golla}, G., {R\"udiger}, G., \& {Wielebinski}, R. 1995, \aap,
  297, 77

\bibitem[{{Elstner} {et~al.}(1990){Elstner}, {Meinel}, \&
  {R{\"u}diger}}]{EMR90}
{Elstner}, D., {Meinel}, R., \& {R{\"u}diger}, G. 1990, Geophys. Astrophys.
  Fluid Dynam., 50, 85, \dodoi{10.1080/03091929008219874}

\bibitem[{{Freidberg}(2014)}]{Freidberg14}
{Freidberg}, J.~P. 2014, {Ideal MHD}, \dodoi{10.1017/CBO9780511795046}

\bibitem[{Garg {et~al.}(2025)Garg, Durrer, \& Schober}]{Garg+25}
Garg, D., Durrer, R., \& Schober, J. 2025.
\newblock \doarXiv{2505.14774}

\bibitem[{{Ghasemi-Nodehi} {et~al.}(2022){Ghasemi-Nodehi}, {Tabatabaei},
  {Sargent}, {Murphy}, {Khosroshahi}, {Beswick}, {Bonaldi}, \&
  {Schinnerer}}]{2022MNRAS.515.1158G}
{Ghasemi-Nodehi}, M., {Tabatabaei}, F.~S., {Sargent}, M., {et~al.} 2022,
  \mnras, 515, 1158, \dodoi{10.1093/mnras/stac1393}

\bibitem[{{Ghosh} {et~al.}(2026){Ghosh}, {Brandenburg}, {Caprini}, {Neronov},
  \& {Vazza}}]{Ghosh+25}
{Ghosh}, O., {Brandenburg}, A., {Caprini}, C., {Neronov}, A., \& {Vazza}, F.
  2026, \prd, 113, 023523, \dodoi{10.1103/4114-mgsp}

\bibitem[{{Ginzburg} \& {Syrovatskii}(1965)}]{GS65}
{Ginzburg}, V.~L., \& {Syrovatskii}, S.~I. 1965, \araa, 3, 297,
  \dodoi{10.1146/annurev.aa.03.090165.001501}

\bibitem[{{Ivanova} \& {Ruzmaikin}(1977)}]{IR77}
{Ivanova}, T.~S., \& {Ruzmaikin}, A.~A. 1977, SvA, 21, 479

\bibitem[{{Jabbari} {et~al.}(2015){Jabbari}, {Brandenburg}, {Kleeorin},
  {Mitra}, \& {Rogachevskii}}]{Jabbari+15}
{Jabbari}, S., {Brandenburg}, A., {Kleeorin}, N., {Mitra}, D., \&
  {Rogachevskii}, I. 2015, \apj, 805, 166, \dodoi{10.1088/0004-637X/805/2/166}

\bibitem[{{Jakab} \& {Brandenburg}(2021)}]{JB21}
{Jakab}, P., \& {Brandenburg}, A. 2021, \aap, 647, A18,
  \dodoi{10.1051/0004-6361/202038564}

\bibitem[{{Kitchatinov} \& {R\"udiger}(1995)}]{Kitchatinov+Ruediger95}
{Kitchatinov}, L.~L., \& {R\"udiger}, G. 1995, \aap, 299, 446

\bibitem[{{Krause} \& {R{\"a}dler}(1980)}]{KR80}
{Krause}, F., \& {R{\"a}dler}, K.-H. 1980, {Mean-Field Magnetohydrodynamics and
  Dynamo Theory} (Oxford: Pergamon Press)

\bibitem[{{Meinel} {et~al.}(1990){Meinel}, {Elstner}, \& {Ruediger}}]{MER90}
{Meinel}, R., {Elstner}, D., \& {Ruediger}, G. 1990, \aap, 236, L33

\bibitem[{{Moffatt}(1978)}]{Mof78}
{Moffatt}, H.~K. 1978, {Magnetic Field Generation in Electrically Conducting
  Fluids} (Cambridge: Cambridge University Press)

\bibitem[{{Murray} {et~al.}(1986){Murray}, {Stanley}, \& {Brown}}]{Murray86}
{Murray}, J.~D., {Stanley}, E.~A., \& {Brown}, D.~L. 1986, Proc. Roy. Soc.
  Lond. Ser. B, 229, 111, \dodoi{10.1098/rspb.1986.0078}

\bibitem[{{Oei}(2025)}]{Oei}
{Oei}, M. S. S.~L. 2025, private information

\bibitem[{{Parker}(1979)}]{Par79}
{Parker}, E.~N. 1979, {Cosmical Magnetic Fields: Their Origin and Their
  Activity} (Oxford: Clarendon Press)

\bibitem[{{Pencil Code Collaboration} {et~al.}(2021){Pencil Code
  Collaboration}, {Brandenburg}, {Johansen}, {Bourdin}, {Dobler}, {Lyra},
  {Rheinhardt}, {Bingert}, {Haugen}, {Mee}, {Gent}, {Babkovskaia}, {Yang},
  {Heinemann}, {Dintrans}, {Mitra}, {Candelaresi}, {Warnecke},
  {K{\"a}pyl{\"a}}, {Schreiber}, {Chatterjee}, {K{\"a}pyl{\"a}}, {Li},
  {Kr{\"u}ger}, {Aarnes}, {Sarson}, {Oishi}, {Schober}, {Plasson}, {Sandin},
  {Karchniwy}, {Rodrigues}, {Hubbard}, {Guerrero}, {Snodin}, {Losada},
  {Pekkil{\"a}}, \& {Qian}}]{PC}
{Pencil Code Collaboration}, {Brandenburg}, A., {Johansen}, A., {et~al.} 2021,
  J. Open Source Software, 6, 2807, \dodoi{10.21105/joss.02807}

\bibitem[{{Peron} {et~al.}(2021){Peron}, {Aharonian}, {Casanova}, {Yang}, \&
  {Zanin}}]{2021ApJ...907L..11P}
{Peron}, G., {Aharonian}, F., {Casanova}, S., {Yang}, R., \& {Zanin}, R. 2021,
  \apjl, 907, L11, \dodoi{10.3847/2041-8213/abcaa9}

\bibitem[{{Perri} {et~al.}(2021){Perri}, {Brun}, {Strugarek}, \&
  {R{\'e}ville}}]{Perri+21}
{Perri}, B., {Brun}, A.~S., {Strugarek}, A., \& {R{\'e}ville}, V. 2021, \apj,
  910, 50, \dodoi{10.3847/1538-4357/abe2ac}

\bibitem[{{Pfrommer} {et~al.}(2022){Pfrommer}, {Werhahn}, {Pakmor},
  {Girichidis}, \& {Simpson}}]{2022MNRAS.515.4229P}
{Pfrommer}, C., {Werhahn}, M., {Pakmor}, R., {Girichidis}, P., \& {Simpson},
  C.~M. 2022, \mnras, 515, 4229, \dodoi{10.1093/mnras/stac1808}

\bibitem[{{Priest}(1982)}]{Priest}
{Priest}, E.~R. 1982, {Solar magnetohydrodynamics.}, Vol.~21 (Cambridge
  University Press)

\bibitem[{{Rempel}(2006)}]{Rempel06}
{Rempel}, M. 2006, \apj, 647, 662, \dodoi{10.1086/505170}

\bibitem[{{Rybicki} \& {Lightman}(1986)}]{Rybicki+Lightman86}
{Rybicki}, G.~B., \& {Lightman}, A.~P. 1986, {Radiative Processes in
  Astrophysics} (Wiley-VCH Verlag GmbH \& Co. KGaA)

\bibitem[{{Sch\"ussler}(1979)}]{Schuessler79}
{Sch\"ussler}, M. 1979, \aap, 72, 348

\bibitem[{Seller \& Sigl(2025)}]{Seller+Sigl2025}
Seller, K., \& Sigl, G. 2025, On the contribution of galaxies to the magnetic
  field in cosmic voids.
\newblock \doarXiv{2510.08025}

\bibitem[{{Steenbeck} \& {Krause}(1969)}]{SK69}
{Steenbeck}, M., \& {Krause}, F. 1969, Astron. Nachr., 291, 49,
  \dodoi{10.1002/asna.19692910201}

\bibitem[{{Stix}(1975)}]{Stix75}
{Stix}, M. 1975, \aap, 42, 85

\bibitem[{{Taziaux} {et~al.}(2025){Taziaux}, {M{\"u}ller}, {Adebahr}, {Basu},
  {Pfrommer}, {Stein}, {Chy{\.z}y}, {Bomans}, {En{\ss}lin}, {Heesen},
  {Kamphuis}, {Soida}, {Wezgowiec}, {Dettmar}, {Das}, \&
  {Tjus}}]{2025A&A...696A.226T}
{Taziaux}, S., {M{\"u}ller}, A., {Adebahr}, B., {et~al.} 2025, \aap, 696, A226,
  \dodoi{10.1051/0004-6361/202453311}

\bibitem[{{Vazza} {et~al.}(2025){Vazza}, {Gheller}, {Zanetti}, {Tsizh},
  {Carretti}, {Mtchedlidze}, \& {Br{\"u}ggen}}]{Vazza+25}
{Vazza}, F., {Gheller}, C., {Zanetti}, F., {et~al.} 2025, \aap, 696, A58,
  \dodoi{10.1051/0004-6361/202451709}

\bibitem[{{Werhahn} {et~al.}(2021){Werhahn}, {Pfrommer}, {Girichidis},
  {Puchwein}, \& {Pakmor}}]{2021MNRAS.505.3273W}
{Werhahn}, M., {Pfrommer}, C., {Girichidis}, P., {Puchwein}, E., \& {Pakmor},
  R. 2021, \mnras, 505, 3273, \dodoi{10.1093/mnras/stab1324}

\bibitem[{{Yang} {et~al.}(1986){Yang}, {Sturrock}, \&
  {Antiochos}}]{yang1986force}
{Yang}, W.~H., {Sturrock}, P.~A., \& {Antiochos}, S.~K. 1986, \apj, 309, 383,
  \dodoi{10.1086/164610}

\end{thebibliography}

\appendix
\section{Dispersion relation for a dynamo with constant coefficients}
\label{DispersionRelation}

\begin{figure}[b]\begin{center}
\includegraphics[width=\columnwidth]{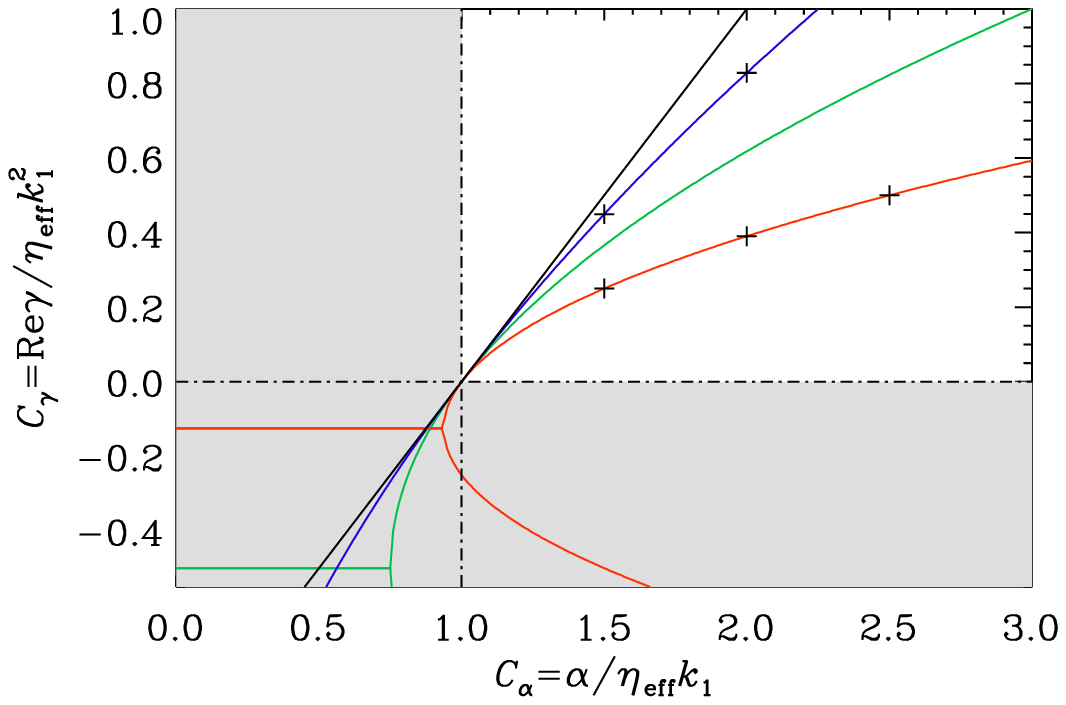}
\end{center}\caption{
Growth rate vs.\ dynamo number $\alpha/\etaT k_1$ for a fixed value of $k=k_1$
and $c/\etaT k=0.5$ (red), 1 (green), and 2 (blue).
The gray areas mark the regions of no growth.
The conventional dependence for $c\to\infty$ corresponds to the black line.
The black plus signs refer to data points obtained using the \textsc{Pencil Code}.
}\label{pcomp}\end{figure}

To illuminate the mean-field dynamo behavior for a finite speed of light and to provide a benchmark for testing purposes,
we discuss here the case of an $\alpha^2$ dynamo with constant coefficients.
Seeking solutions proportional to $\exp(\gamma t+\ii\kk\cdot\xx)$,
the dispersion relation $\gamma(\kk)$ can be obtained by solving for the roots of
\begin{equation}
\left[\gamma+\etaT k^2\left(1+\frac{\gamma^2}{c^2k^2}\right)\right]^2-\alpha^2 k^2=0.
\end{equation}
In the limit $c\to\infty$, the dispersion relation agrees with the conventional one:
\begin{equation}
\gamma=\pm|\alpha|k-\etaT k^2.
\end{equation}
\FFig{pcomp} shows $\gamma(\alpha)$ for a fixed value of $k=k_1$ and three values of $c/\etaT k_1$.
The dependence matches the conventional one for $c\to\infty$.
We see that the effect of a finite speed of light is to lower the value of $\gamma$.

\begin{figure}[t]\begin{center}
\includegraphics[width=\columnwidth]{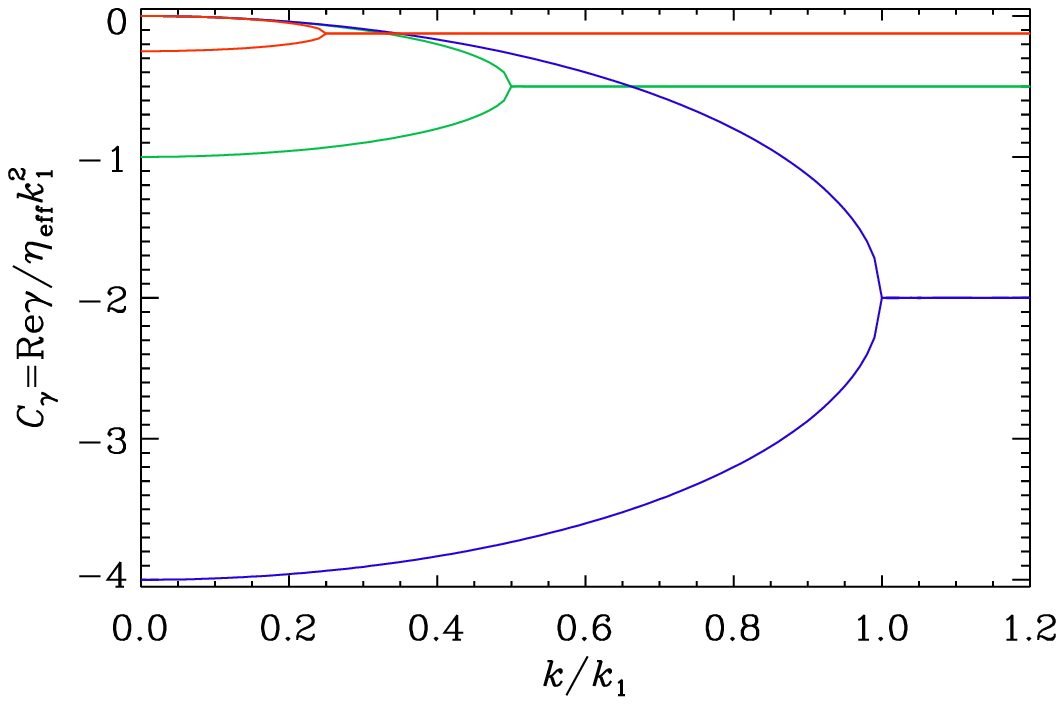}
\end{center}\caption{
Normalized growth rate vs.\ $k/k_1$ for $\alpha=0$ and $c/\etaT k_1=0.5$ (red), 1 (green), and 2 (blue).
For $k>c/2\etaT$, the two solutions are oscillatory with a complex conjugated pair of eigenvalues whose real part is independent of $k$.
}\label{pcomp_alp0}\end{figure}

For $\alpha=0$, the growth rate is always negative.
There are then two branches, given by
\begin{equation}
\gamma_\pm=-\left[1\pm\sqrt{1-(2\etaT k/c)^2}\right]\,c^2/2\etaT.
\end{equation}
For $k>c/2\etaT$, i.e., for $k/k_1>c/2\etaT k_1$, $\gamma$ is complex, i.e., the solutions are oscillatory
with the growth rate $\Rey\gamma=-c^2/2\etaT$, which is independent of $k$.
This is shown in \Fig{pcomp_alp0}, where we plot the negative normalized growth rate versus $k/k_1$ for different values of $c/\etaT k_1$.
Evidently, only for $c/\etaT k_1>2$ are the solutions nonoscillatory for all wavenumbers.

\begin{figure}[t]\begin{center}
\includegraphics[width=\columnwidth]{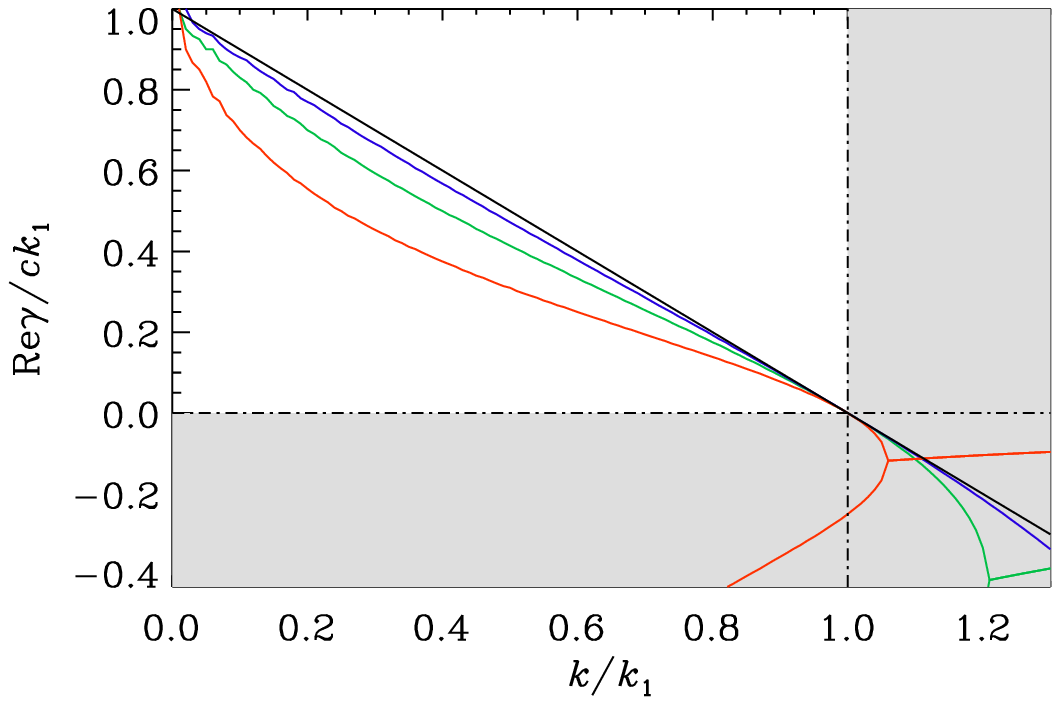}
\end{center}\caption{
Front speed $\gamma/k$ (in units of $c$) versus $k$ (in units of $c/\etaT$) for $\alpha/c=1$ and
$c/\etaT k=0.5$ (red), 1 (green), and 2 (blue).
As in \Fig{pcomp}, the gray areas mark the regions of no growth.
The conventional line for $c\to\infty$ corresponds to the black line.
}\label{pcomp_speed}\end{figure}

For an exponentially growing magnetic field, the spatial spreading can be
described by writing $\meanBB\propto e^{-\kappa(z-c_\mathrm{front}t)}$,
where $c_\mathrm{front}$ is the front speed, $\kappa
c_\mathrm{front}=\gamma$ is the growth rate, and $\kappa=k$.
This relation is only valid outside the dynamo-active region, where $\alpha=0$.
Since the effect of a finite speed of light is to lower the value of $\gamma$,
this also lowers the front speed $\gamma/k$.
For the conventional dispersion relation, the front speed is given by
\begin{equation}
c_\mathrm{front}=\gamma/k=|\alpha|-\etaT k.
\end{equation}
For finite values of $c$, the front speed is lowered,
except for the marginal point when $k$ exceeds the critical value below which dynamo action is possible.
This is shown in \Fig{pcomp_speed} for three finite values of $c/\etaT k$.
We see that the line for $c/\etaT k=2$ is already rather close to the case $c\to\infty$.
As discussed in the bulk of the present paper, $c/\etaT k$ is indeed very large
for realistic values of $\etaT$ and $k$.
Therefore, the Faraday displacement current can indeed be neglected for all practical purposes.

As expected, the front speed never exceeds the speed of light.
By contrast, however, the input parameters $\alpha$ and $\etaT k$,
which also have the dimensions of a speed, can exceed the value of $c$.
Whether such values can be physically realized is, however, open to questions.

\section{Power-law growth of $\etaT$}
\label{PowerLawGrowth}

\begin{figure}\begin{center}
\includegraphics[width=\columnwidth]{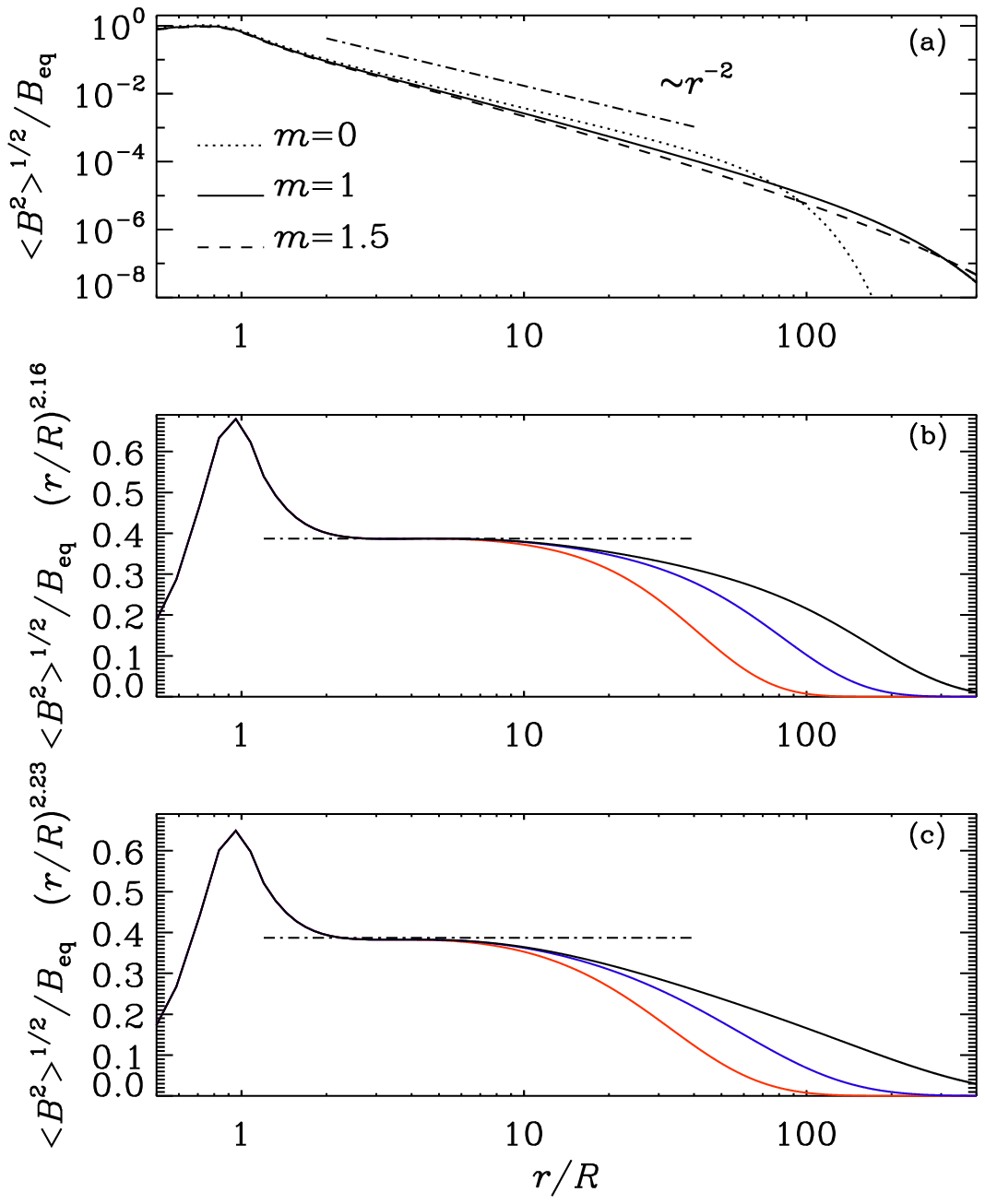}
\end{center}\caption{
(a) Radial magnetic field profiles for the quadrupolar Runs~G, G', and G''.
Runs~G' and G'' have a radially increasing $\etaT(r)$ profile.
In panels (b) and (c), the mean magnetic field profiles are compensated by $r^{2.16}$ and $r^{2.23}$, respectively.
The red, blue, and black lines correspond to times
$t/\tau_\mathrm{diff}=100$, 250, and 600 for Run~G', and
$t/\tau_\mathrm{diff}=50$, 100, and 250 for Run~G''.
}\label{pprof_comp_RunG}\end{figure}

For the quadrupolar case, the inverse-quadratic radial falloff is still approximately obeyed even when $\etaT$ is not constant for $r>R$.
This is demonstrated in \Fig{pprof_comp_RunG}, where we compare the mean radial magnetic field profiles for Runs~G, G', and G''.
The slopes do not depart markedly from the $r^{-2}$ decay.
For a more quantitative demonstration, we also show, for Runs~G' and G'', the profiles
compensated by $r^{+n}$ with $n=2.16$ and with $n=2.23$, respectively.
This confirms that the decay exponents given in \Tab{TSummary} are accurately determined.

\section{Derivation of Equation~(41)}
\label{DerivationEq39}

The purpose of this appendix is to provide the detailed derivation of \Eq{Eq39}.
Following \cite{Rybicki+Lightman86}, we consider a power-law distribution of cosmic rays such as $N(E) d E=\mathrm{C} E^{-\ggamma} d E, \quad E_{\rm min}<E<E_{\rm max}$
with a particle distribution index $\ggamma$ corresponding to a spectral index $s$ given by $s=(\ggamma-1)/2$, such that $P_{\rm tot}(\omega) \propto \omega^{-(\ggamma-1) / 2}$.
The normalization can be determined from the cosmic-ray number density
\begin{equation}
    n_{\mathrm{CR}}=C \frac{E_{\max }^{1-\ggamma}-E_{\min }^{1-\ggamma}}{1-\ggamma},
\end{equation}
for $E_{\rm min}<E<E_{\rm max}$ known as the band-limited case.
However, if $E_{\rm max} \rightarrow \infty$, normalization $C$ depends only on the threshold energy $E_{\rm min}$ for $\ggamma > 1$ as
\begin{equation}
C=n_{\mathrm{CR}}(\ggamma-1) E_{\min }^{\ggamma-1}.
\end{equation}
The total synchrotron power per unit volume per unit frequency irradiated by such a population of cosmic rays can be written as
\begin{eqnarray}
P_{\mathrm{tot}}(\nu) 
&=\frac{\sqrt{3}\, q^3 \mathrm{C}\, B \sin \alpha}{m c^2 (\ggamma+1)} \,C_2\,
\left(\frac{2 \pi m c \nu}{3 q B \sin \alpha}\right)^{\frac{1-\ggamma}{2}},
\end{eqnarray}

Based on this, the optically thick surface brightness can be expressed in general as
\begin{equation}
I_\nu \propto n_{\mathrm{C} R}\left(>E_{\min }\right) E_{\min }^{\ggamma-1} B^{(1+\ggamma)/2} L \nu^{(1-\ggamma)/2}.
\end{equation}
Taking the equipartition magnetic field $B = B_{\rm eq}$, 
averaging over the pitch-angle factor $ \langle \sin^{(1 + \ggamma)/2} \alpha \rangle$, and finally setting the index $\ggamma = 3$, we obtain an optically thick surface brightness of the form
\begin{eqnarray}
I_\nu 
&\approx 5 \times 10^{-6} \,\mathrm{Jy}/\mathrm{sr}
\left(\frac{n_{\mathrm{CR}}}{10^{-7}\,\mathrm{cm}^{-3}}\right)
\left(\frac{E_{\min}}{100\,\mathrm{MeV}}\right)^2 \nonumber \\
&\times 
\left(\frac{B}{1\,\mu\mathrm{G}}\right)^2
\left(\frac{L}{1\,\mathrm{kpc}}\right)
\left(\frac{\nu}{150\,\mathrm{MHz}}\right)^{-1}.
\end{eqnarray}
Replacing $(\nu/150\,\mathrm{MHz})^{-1}$ with $(\lambda/2\,\mathrm{m})$ and using
$E_{\min}=100\,\mathrm{MeV}$, we arrive at \Eq{Eq39}.

\end{document}